\documentclass[12pt]{iopart}

\usepackage{amsmath, amsfonts, amssymb, amsthm}
\usepackage{physics, empheq}
\usepackage{hyperref}
\usepackage{graphicx}
\usepackage[svgnames]{xcolor}
\usepackage{siunitx}  % For values with units
\usepackage{tcolorbox}
\begin{document}

\title{Minimal model of quasi-cyclic behaviour in turbulence driven by Taylor--Green forcing}

\author{Ryo Araki}
\address{
  Univ Lyon, \'Ecole Centrale de Lyon, CNRS, Univ Claude Bernard Lyon 1, INSA Lyon,
  LMFA, UMR5509, 69130, \'Ecully, France
}
\address{
  Graduate School of Engineering Science,
  Osaka University,
  1-3 Machikaneyama, Toyonaka, Osaka 560-8531, Japan
}
\ead{araki.ryo@ec-lyon.fr}

\author{Wouter J.~T.~Bos}
\address{
  Univ Lyon, \'Ecole Centrale de Lyon, CNRS, Univ Claude Bernard Lyon 1, INSA Lyon,
  LMFA, UMR5509, 69130, \'Ecully, France
}
\author{Susumu Goto}
\address{
  Graduate School of Engineering Science,
  Osaka University,
  1-3 Machikaneyama, Toyonaka, Osaka 560-8531, Japan
}

\vspace{10pt}
\begin{indented}
\item[] March 2023
\end{indented}

\begin{abstract}
  We attempt to formulate the simplest possible model mimicking turbulent dynamics, such as quasi-cyclic behaviour (QCB), using only three variables.
  To this end, we first conduct direct numerical simulations of three-dimensional flow driven by the steady Taylor--Green forcing to find a similarity between a stable periodic orbit (SPO) at a small Reynolds number (\(\Re\)) and turbulent QCB at higher \(\Re\).
  A close examination of the SPO allows the heuristic formulation of a three-equation model, representing the evolution of Fourier modes in three distinct scales.
  The model reproduces the continuous bifurcation from SPO to turbulence with QCB when \(\Re\) is varied.
  We also demonstrate that, by changing model parameters, the proposed model exhibits a discontinuous transition from steady to chaotic solutions without going through an SPO.
\end{abstract}

%\maketitle

% ==================================================
\section{Introduction}
\label{sec:Introduction}
% ==================================================

% ++++++++++++++++++++++++++++++++++++++++++++++++++
%\subsection{List of previous research}
\label{subsec:List of previous research}
% ++++++++++++++++++++++++++++++++++++++++++++++++++

The dynamics of turbulent flows is determined by the collective behaviour of a large number of interacting modes.
The very large number of triadic interactions between these modes, even in moderately turbulent flows, prevent us from understanding the global flow features through the direct analysis of the interactions~\cite{Kraichnan1971_inertial, Domaradzki1990_local, Waleffe1992_the_nature}.
To gain such understanding, the complexity of the description needs to be drastically reduced.

Systematic approaches for that purpose have been applied to turbulence research, such as proper orthogonal decomposition (POD)~\cite{Bakewell1967_viscous, Berkooz1993_the_proper, Holmes1996_turbulence} or Galerkin truncation~\cite{Rempfer2000_on_low-dimensional}.
For a list of the key publications in projection-based reduced-order modelling, see a recent review article~\cite{Ahmed2021_on_closures}.
In these approaches, the complex dynamics are dissected by projecting the Navier--Stokes equations on a low-dimensional basis of eigenfunctions.

A more heuristic approach to reduce complexity is the direct modelling of the dynamics by a small number of variables retaining a number of constraints (such as energy or helicity conservation).
A well-known example of such an approach is the development of shell models~\cite{Biferale2003_shell, Ditlevsen2010_turbulence}, as first proposed by Obukhov~\cite{Obukhov1971_some}.
This approach bypasses the definition of the basis function by directly modelling the dynamics of an ensemble of modes.

In the current study, we combine observations of direct numerical simulations (DNS) and heuristic modelling.
We assess the detailed dynamics of a numerical simulation and investigate the interactions between small groups of Fourier modes.
Inspired by the form of the Navier--Stokes equations, we represent the full dynamics by an ODE system of three interacting variables, yielding a sort of shell model with both dyadic and triadic interactions between the groups of modes.

The turbulent flow we characterise is incompressible Navier--Stokes turbulence, driven by a large-scale steady forcing in a spatially periodic domain.
% Such flow is shown to exhibit temporally recurrent large-scale fluctuations.
At low Reynolds numbers, the considered flow becomes temporally periodic.
We will show that, even for this specific periodic flow, retaining the modes governing both energy and enstrophy in the flow considered in the present investigation leads to a subset of several dozens of complex-valued Fourier modes.
This results in a system of an important number of coupled ODE, which will not allow analytical treatment.
Therefore, using a more heuristic approach, we analyse the periodic flow, identify the key interactions between scales, and formulate the simplest model which retains these interactions and the forcing and dissipation mechanisms.
This approach allows us to formulate a model containing only three degrees of freedom, reproducing certain characteristics of the investigated fluid flow.
In particular, one feature we want to reproduce with our model is quasi-cyclic behaviour (QCB).

A number of laminar and turbulent flows display QCB.
An illustrative example is vortex shedding behind an obstacle.
For low Reynolds number (\(\Re\)), the so-called von K\'arm\'an vortex street behind a cylinder is perfectly periodic, which corresponds to a stable periodic solution (SPO) in phase space.
Even when the flow becomes fully turbulent at higher \(\Re\), this periodicity is still present, though the stochastic nature of turbulence motion prevents the system from being perfectly periodic.
This close-to-periodic motion, embedded in turbulent fluctuations, is what we will call QCB.

Another important example of QCB is the temporal behaviour of turbulent channel flow, where a self-sustaining process governs the dynamics~\cite{Waleffe1995_hydrodynamic, Hamilton1995_regeneration, Panton2001_overview}.
In particular, in small channel flow domains (the so-called minimal flow unit), close to periodic behaviour is observed (Fig.~6 of Ref.~\cite{Jimenez1991_minimal}).
The simplified descriptions of this phenomenon are specific to channel flow or the simplified case of Waleffe flow~\cite{Waleffe1997_on_a_self-sustaining, Thomas2014_self-sustaining, Thomas2015_a_minimal, Alizard2019_restricted, Cavalieri2021_structure}.
Non-trivial QCB was also observed~\cite{Araki2021_quasiperiodic} in a confined cylindrical flow between two counter-rotating disks (the so-called von K\'arm\'an flow).
Furthermore, QCB is observed in periodic box flow with steady forcing~\cite{Yasuda2014_quasi-cyclic, Goto2017_hierarchy, vanVeen2018_transitions}.
These observations suggest that such dynamics might be more general than wall-bounded flow or flow behind obstacles.
Moreover, a vast amount of recent research is dedicated to identifying unstable periodic orbits (UPO) embedded in turbulent flows~\cite{Auerbach1987_exploring, Kawahara2001_periodic, Toh2003_periodic, Kawasaki2005_statistics, Kawahara2012_significance, Lucas2017_sustaining, vanVeen2019_time}.
Recently, the periodic orbit-aided reduced-order model was discussed~\cite{Yalniz2021_coarse}.
% These studies show that UPOs reproduce statistics and dynamical properties of turbulence.

In the present study, we construct a three-mode model which exhibits turbulent QCB.
Our strategy is as follows.
First, we conduct DNS of turbulence in a periodic cube to find an SPO at low \(\Re\), which resembles the turbulent QCB observed at higher \(\Re\) (see \S~\ref{sec:Observation of Quasi-cyclic behaviour}).
Secondly, in \S~\ref{sec:Three-equation model}, we construct a minimal model by carefully examining nonlinear interactions in the SPO.
Then, in \S~\ref{subsec:Bifurcation from SPO to chaos with QCB}, we demonstrate that an SPO of the model bifurcates to a chaotic solution that indeed shows QCB.
In \S~\ref{subsec:Subcritical bifurcation to chaos}, we also demonstrate that another route, via subcritical transition, can be possible for another set of model parameters to show the potential ability of the proposed model to be extended to other types of turbulence.

% ==================================================
\section{Observation of Quasi-cyclic behaviour}
\label{sec:Observation of Quasi-cyclic behaviour}
% ==================================================

To illustrate the features we want to reproduce and guide the formulation of a minimal model reproducing these features, we conduct numerical simulations of both turbulent and temporally periodic flows with the same type of forcing.
More precisely, we conduct DNS of three-dimensional incompressible flow governed by the Navier--Stokes equations,
\begin{equation}
  \pdv{\vb*{u}}{t}
    + \qty(\vb*{u} \vdot \grad) \vb*{u}
    = -\grad{p}
    + \nu \laplacian \vb*{u}
    + \vb*{f},
  \label{eq:Navier--Stokes}
\end{equation}
with a steady forcing of the two-dimensional Taylor--Green type~\cite{Yasuda2014_quasi-cyclic, Goto2017_hierarchy, vanVeen2018_transitions},
\begin{equation}
  \vb*{f}
    = \qty(- f_0 \sin x \cos y, f_0 \cos x \sin y, 0),
    % &= \qty(- f_0 \sin \frac{2\pi k_f^{(x)}}{L} x \cos \frac{2\pi k_f^{(y)}}{L} y, f_0 \cos \frac{2\pi k_f^{(x)}}{L} x \sin \frac{2\pi k_f^{(y)}}{L} y, 0),
  \label{eq:body_force}
\end{equation}
and the continuity equation, \(\div{\vb*{u}} = 0\).
Here, \(\vb*{u}\), \(p\), and \(\vb*{f}\) are the velocity, pressure, and forcing fields, respectively.
The forcing amplitude \(f_0\) is set to unity.
The only control parameter is the kinematic viscosity \(\nu\).
We employ a pseudo-spectral method in a \((2\pi)^3\) periodic box.
See~\ref{sup:Direct Numerical Simulations} for details of the DNS.
We define the Reynolds number and the characteristic timescale of large-scale flow as
\begin{equation}
  \Re \equiv \frac{\sqrt{f_0}}{\abs{\vb*{k}_f}^{3/2} \nu} \qand
  T \equiv \frac{1}{\sqrt{\abs{\vb*{k}_f} f_0}} = 0.840,
  \label{eq:def_Re_T}
\end{equation}
respectively.
Here, \(\vb*{k}_f = (\pm 1,\pm 1,0)\) is the wavevector of the forcing~\eqref{eq:body_force}.

\begin{figure}
  \centering
  \includegraphics[width=0.8\textwidth]{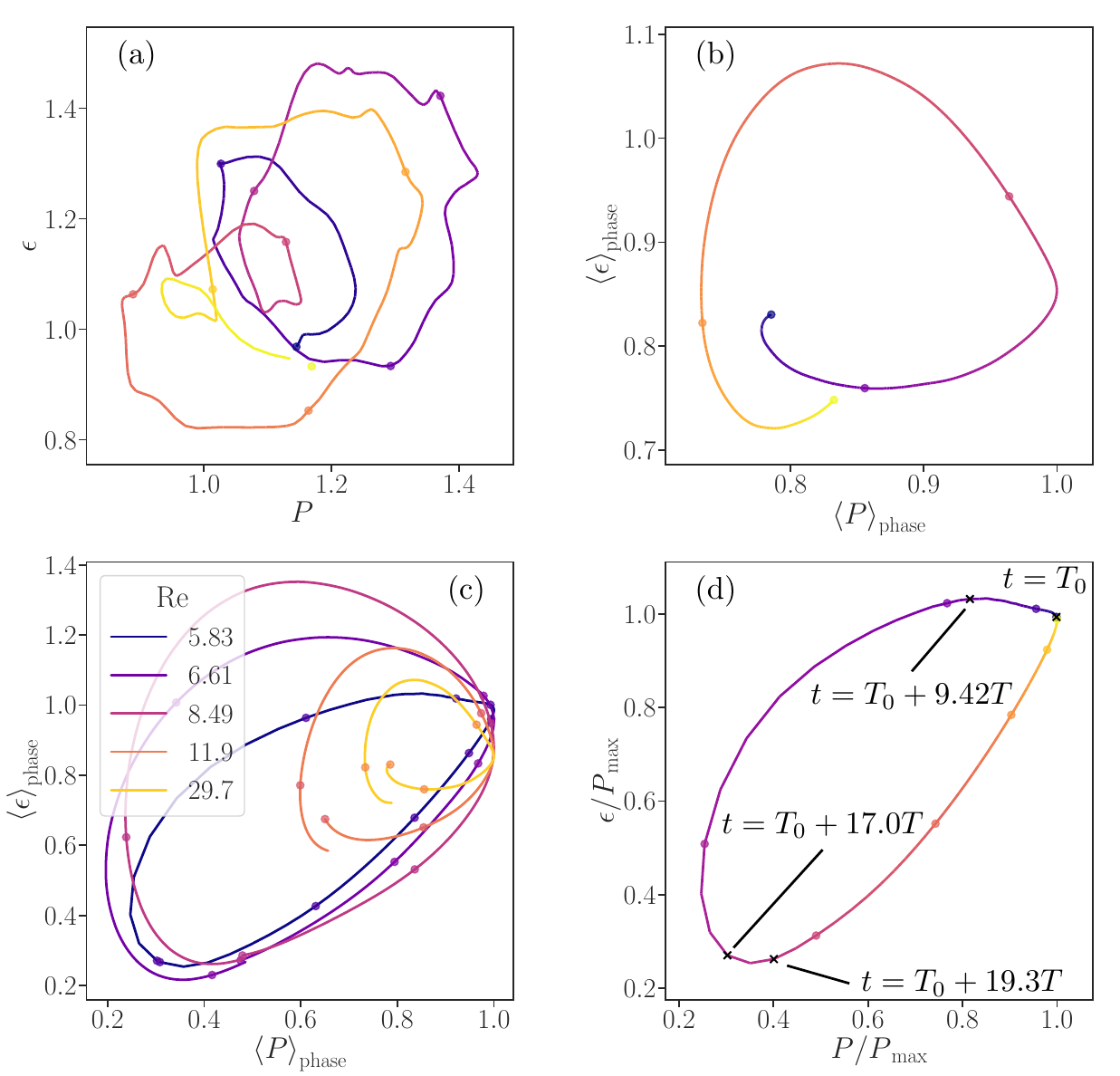}
  \caption{
    (a) Parametric plots of the instantaneous values of the energy dissipation rate \(\epsilon(t)\) and the energy input rate \(P(t)\) for the turbulent flow at \(\Re = 29.7\) for \(50 T\) [See Fig.~\ref{fig:turbulence_snapshot_timeseries}~(b) in~\ref{sup:Detailed procedure of the phase averaging}].
    (b) Phase-averaged values \(\expval{P}_\mathrm{phase}\) and \(\expval{\epsilon}_\mathrm{phase}\) at \(\Re = 29.7\) [same as in (a)] for \(20T\).
    (c) Parametric plots of the phase averaged values \(\expval{P}_\mathrm{phase}\) and \(\expval{\epsilon}_\mathrm{phase}\) at four different values of \(\Re\) (\(29.7\), \(11.9\), \(8.49\), and \(6.61\)).
    Note that the orbit in panel (b) at \(\Re = 29.7\) is re-plotted in panel (c).
    (d) The SPO at \(\Re = 5.83\) is shown with a coloured line.
    Note that this orbit is also shown in panel (c).
    The gap between two consecutive dots corresponds to \(5T\) for all panels.
    In panels (a), (b), and (d), the time evolves from dark to light colours.
    Four black cross symbols in panel (d) denote the instances shown in Fig.~\ref{fig:periodic_snapshots}.
  }
  \label{fig:QCB_turbulence}
\end{figure}

Figure~\ref{fig:QCB_turbulence} shows the temporal evolution of the energy input rate \(P(t) \equiv \expval{\vb*{f} \vdot \vb*{u}}\) against the energy dissipation rate \(\epsilon(t)\) given by \(\nu \expval{\abs{\vb*{\omega}}^2}\) for various Reynolds numbers.
Here, \(\expval{\cdot}\) denotes the spatial average and \(\vb*{\omega} \equiv \grad \times \vb*{u}\).

Figure~\ref{fig:QCB_turbulence}~(a) shows the turbulent time series, where the time-averaged Taylor scale-based Reynolds number \(\expval{\Re_\lambda}_t\) is about \(90\).
Note that \(\expval{\cdot}_t\) denotes the time average.
We also show the snapshot of this flow in Fig.~\ref{fig:turbulence_snapshot_timeseries}~(a) in~\ref{sup:Detailed procedure of the phase averaging}.
The time series exhibits QCB in a counter-clockwise direction behind the chaotic fluctuations.
This time delay between the large- and small-scale representatives (i.e. \(P(t)\) and \(\epsilon(t)\)) reflects the causal nature of the energy cascade.

We apply a phase average to the complex time series of \(P(t)\) and \(\epsilon(t)\) conditioned on the local maxima of \(P(t)\) in order to extract smooth, time-delayed oscillations shown in Fig.~\ref{fig:QCB_turbulence}~(b).
See~\ref{sup:Detailed procedure of the phase averaging} for the detailed procedure.
We denote the phase-averaged quantities by \(\expval{\cdot}_\mathrm{phase}\).
These results suggest that the QCB of turbulent flow driven by the steady body force~\eqref{eq:body_force} is robust.
Such QCB is also shown in Fig.~12 of Ref.~\cite{Goto2017_hierarchy} for two different forcing types at even higher \(\Re\).
The physical origin of QCB is rooted in the energy cascading process from larger to smaller scales.
Since the coherent structures at these scales are composed of a large number of Fourier modes, to describe the QCB in terms of Fourier modes, we need to understand the underlying nonlinear interactions among them.
However, identifying the direct cause of QCB from tens of thousands of excited Fourier modes seems illusory.
Thus, we decrease \(\Re\) to reduce the complexity of the flow.

In Fig.~\ref{fig:QCB_turbulence}~(c), we show the phase-averaged plots of the parametric time series of \(P(t)\) and \(\epsilon(t)\) for four different values of \(\Re\).
The change in the shape of the parametric plots is gradual, suggesting that the quasi-cyclic orbit in the turbulent flow is continuously connected to an SPO at \(\Re \approx 5.83\), which is also shown in Fig.~\ref{fig:QCB_turbulence}~(d) for comparison.
As will be shown in Fig.~\ref{fig:periodic_snapshots} below, the SPO at \(\Re = 5.83\) is not the laminar solution of the system which corresponds to a purely two-dimensional structure resulting from a balance between viscous stress and the forcing~\eqref{eq:body_force}.
We emphasise that this SPO plays a key role in constructing our model.

We find that the amplitude and the period of the periodic and quasi-cyclic flows monotonically increase when we decrease \(\Re\) from \(29.7\) to \(5.83\).
This does not prove that the dynamics are identical, but the turbulent QCB and periodic flow seem to share the same driving mechanism.
Note that in high-Reynolds-number turbulence beyond \(\Re = 30\), the amplitude and period seem to saturate to values of the same order as in Fig.~\ref{fig:QCB_turbulence}~(b) (See Fig.~12 of Ref.~\cite{Goto2017_hierarchy}).

\begin{figure}
  \centering
  \includegraphics[width=0.7\textwidth]{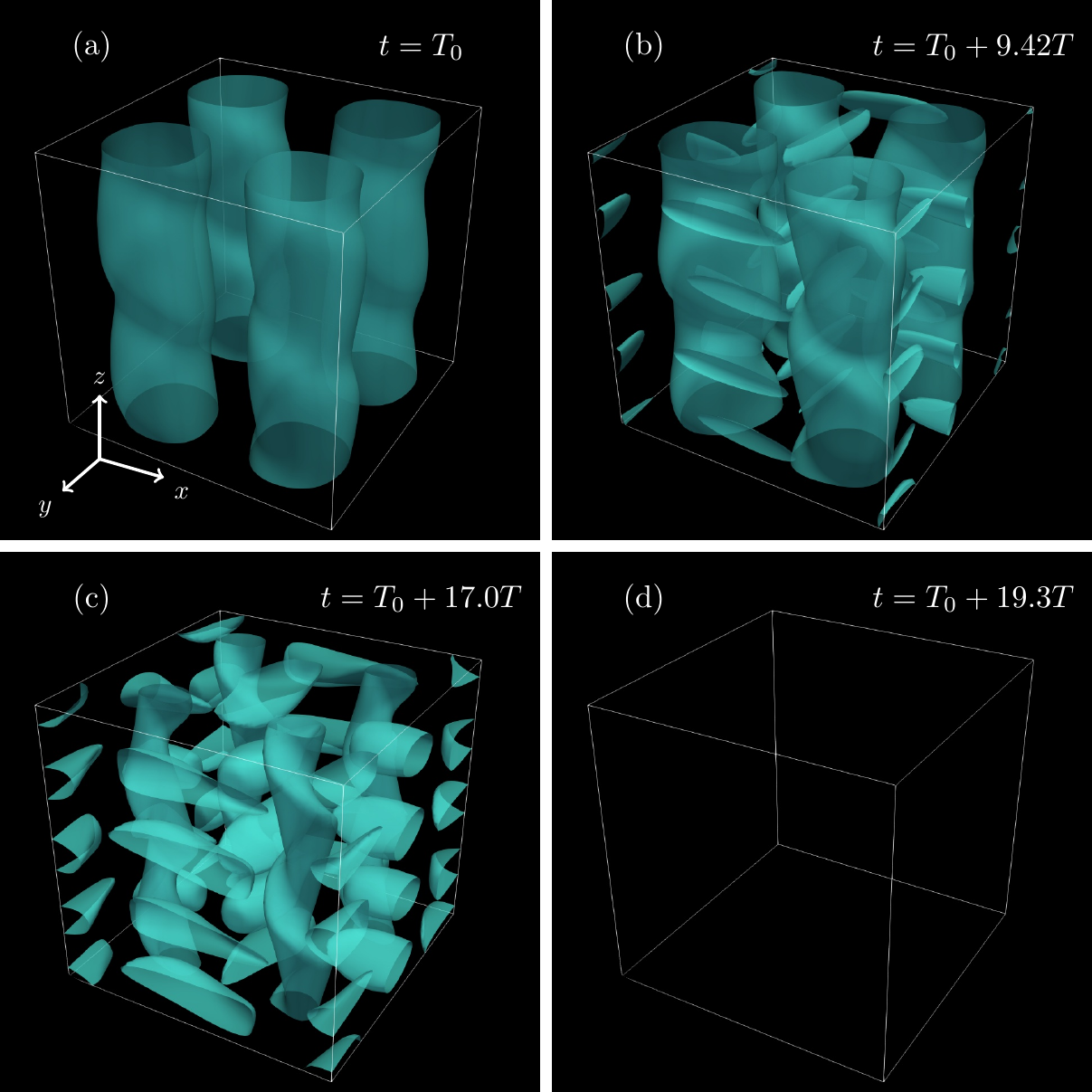}
  \caption{
    Visualisation of vortical structures of the SPO (at \(\Re = 5.83\)) at four instances with isosurface of \(\abs{\vb*{\omega}} = 5\).
    See Fig.~\ref{fig:QCB_turbulence}~(d) for the corresponding instances.
  }
  \label{fig:periodic_snapshots}
\end{figure}

In Fig.~\ref{fig:periodic_snapshots}, we visualise the periodic flow (i.e. SPO) discussed in Fig.~\ref{fig:QCB_turbulence}~(d), similar to the three-dimensional periodic solution reported in~\cite[Fig.~5]{vanVeen2018_transitions}.
We distinguish four large-scale columnar vortices associated with the Taylor--Green force~\eqref{eq:body_force} and counter-rotating pairs of smaller vortices perpendicular to them.
Note that we do not perform low-pass filtering [See Fig.~\ref{fig:turbulence_snapshot_timeseries}~(a) of~\ref{sup:Detailed procedure of the phase averaging}] since there is no significant scale separation in the periodic flow.
Nevertheless, we can observe a one-step energy cascading process from the four large-scale columnar vortices to smaller-scale lateral vortices.
More concretely, we observe only large-scale vortices at \(t = T_0\) [Fig.~\ref{fig:periodic_snapshots}~(a)], then the energy cascade starts to create smaller-scale vortices [Fig.~\ref{fig:periodic_snapshots}~(b), \(t = T_0 + 9.42 T\)], while the large-scale vortices get weaker [Fig.~\ref{fig:periodic_snapshots}~(c), \(t = T_0 + 17.0 T\)].
Afterwards, the energy dissipation dominates to weaken smaller-scale vortices, and then the entire system becomes calm [Fig.~\ref{fig:periodic_snapshots}~(d), \(t = T_0 + 19.3 T\)].
When small-scale vortices disappear, energy input by the external force exceeds dissipation to reestablish the large-scale vortices, and the system returns to the initial state [Fig.~\ref{fig:periodic_snapshots}~(a)].
We emphasise that this periodic behaviour is similar to turbulent QPB observed at higher \(\Re\) (Fig.~\ref{fig:turbulence_snapshot_timeseries} of~\ref{sup:Detailed procedure of the phase averaging} and Figs.~12-17 of Ref.~\cite{Goto2017_hierarchy}).
This similarity manifests itself in the continuous change between the SPO and turbulence seen in Fig.~\ref{fig:QCB_turbulence}.

In the next section, we analyse the SPO to unveil the essential physics behind QCB.
Even though we have not rigorously shown the connection between the SPO and turbulence, we hope to obtain new insights into QCB in Navier--Stokes flow by dissecting the SPO.

% ==================================================
\section{Three-equation model}
\label{sec:Three-equation model}
% ==================================================

% ++++++++++++++++++++++++++++++++++++++++++++++++++
\subsection{Construction of the model}
\label{subsec:Construction of the model}
% ++++++++++++++++++++++++++++++++++++++++++++++++++

Our objective is to construct the simplest possible model capable of reproducing QCB while retaining a close connection with the structure of the Navier--Stokes equations~\eqref{eq:Navier--Stokes}.
For this purpose, we recall that in a Fourier representation of~\eqref{eq:Navier--Stokes}, the individual modes \(q_i\) for the \(i\)th wavevector \(\vb*{k}_i\) are governed by~\cite{Kraichnan1958_irreversible, Kraichnan1988_reduced},
\begin{equation}
  \qty(\pdv{t} + \nu \abs{\vb*{k}_i}^2) q_i = \sum_{j,m} A_{ijm} q_j q_m + f_i,
  \label{eq:NS_mode}
\end{equation}
where \(f_i\) is the forcing applied to the \(i\)th mode, and \(A_{ijm}\) are the coupling constants resulting from the advection and pressure terms of~\eqref{eq:Navier--Stokes}.
The nonlinear term associated with triad interactions rapidly yields an overwhelming complexity when the number of retained modes increases.
Even in our SPO, a large number of modes are dynamically active. In order to develop an analytically tractable model, we use a coarse-graining approach where we group subsets of Fourier modes and represent each group by a single variable, leading to a sort of shell-model~\cite{Obukhov1971_some,Ditlevsen2010_turbulence}.

\begin{figure}
  \centering
  \includegraphics[width=0.6\textwidth]{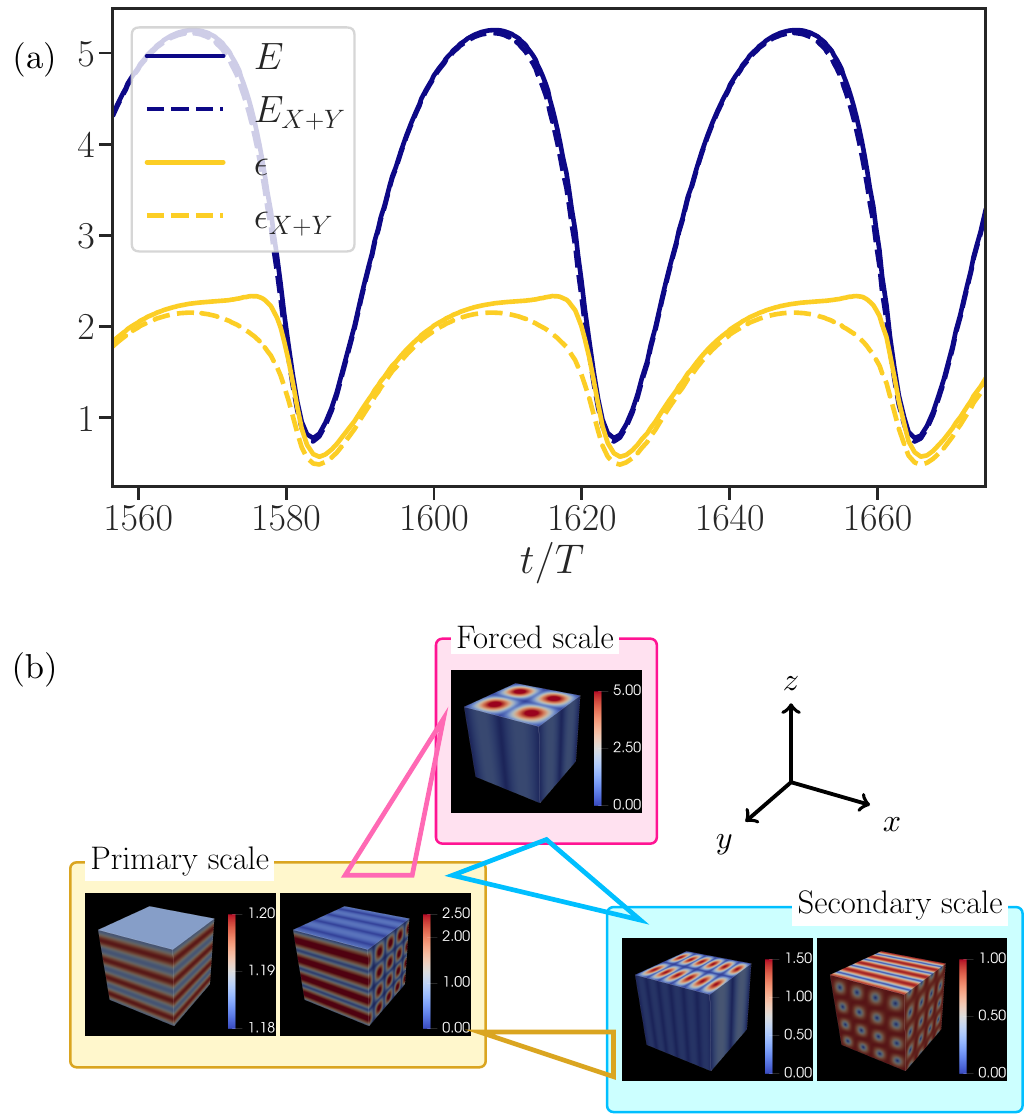}
  \caption{
    (a) Time series of energy \(E(t)\) and energy dissipation rate \(\epsilon(t)\) computed from all modes (solid lines) and those of the forced plus the primary energetic modes, denoted by \((\cdot)_{X+Y}\) (dashed lines).
    (b) Schematic of three different scales: ``forced'', ``primary'', and ``secondary''.
    We visualise \(\abs{\vb*{\omega}}\) distributions of typical Fourier modes in each scale.
    The forced scale corresponds to \(\vb*{k}_f = (\pm 1,\pm 1,0)\).
    In the primary scale, we visualise \(\vb*{k} = (0,0,\pm 2)\) and \((0,\pm 2,\pm 2)\) modes.
    % Refer Fig.~\ref{fig:triad_interactions_3DPF} for the full list of the primary scale.
    For the secondary scale, we visualise \(\vb*{k} = (\pm 3,\pm 1,0)\) and \((\pm 2,\pm 2,\pm 2)\) modes for example.
    Note that the contributions of modes with all the possible sign combinations \((\pm k_x, \pm k_y, \pm k_z)\) are gathered in the visualisations.
    Triangles denote triad interactions between different scales.
    For details of interactions between the forced and primary scales, see Fig.~\ref{fig:triad_interactions_3DPF} in \ref{sup:Primary energetic modes of the periodic flow}.
  }
  \label{fig:three_scales_schematic}
\end{figure}

The shells or groups used in our model are not regrouping modes as a function of scale using a rigorous criterion but as a function of the type of nonlinear interactions and energetic content.
Indeed, we investigate the Fourier decomposition of the SPO to find that only Fourier modes with wavevectors \((k_x, k_y, k_z)\) of
\begin{equation}
  (\pm 1, \pm 1,     0),\>  % (= \vb*{k}_f), \>
  (\pm 1,     0,     0), \>
  (0, \pm 1,     0), \>
  (0,     0, \pm 2), \>
  (\pm 1, \pm 1, \pm 2), \>
  (\pm 2,     0, \pm 2), \> \text{and} \>
  (0, \pm 2, \pm 2)
  \label{eq:seven_primal_modes}
\end{equation}
are responsible for \SI{98}{\percent} of its energy.
See~\ref{sup:Primary energetic modes of the periodic flow} for details of these energetic modes.
Figure~\ref{fig:three_scales_schematic}~(a) illustrates that the time evolution of the kinetic energy is closely reproduced, retaining only these modes.

A close inspection of the seven modes shows that all the nonlinear interactions involve the forced mode and two of the six other modes (See Fig.~\ref{fig:triad_interactions_3DPF} in \ref{sup:Primary energetic modes of the periodic flow}).
In the following, \(X \in \mathbb{R}\) denotes the characteristic velocity of the forced modes \(\vb*{k} = (\pm 1, \pm 1,     0)\) and \(Y \in \mathbb{R}\) corresponds to that of the remaining modes in~\eqref{eq:seven_primal_modes}.
At this point, we suppose that there are only these two classes of modes and that we represent each class by a single, real variable.
Furthermore, we assume~\eqref{eq:NS_mode} to govern the interaction of these two variables, \(X\) and \(Y\), yielding,
\begin{equation}
\begin{aligned}
  \dv{X}{t}
      &= - A Y^2 &
      &- \nu K_X^2 X
      + F, \\
  \dv{Y}{t}
    &= + A X Y &
    &- \nu K_Y^2 Y,
  \label{eq:LV_like_model}
\end{aligned}
\end{equation}
with a coefficient \(A>0\), typical wavenumbers \(K_\alpha > 0\) with \(\alpha \in \qty{X, Y}\), and a steady force \(F > 0\).
The first term on the RHS of each equation represents the nonlinear coupling between $X$ and $Y$. This interaction conserves the global energy, \((X^2+Y^2)/2\).
Note that since we model the triadic nonlinear term of~\eqref{eq:NS_mode} by regrouping the modes into two families (See Fig.~\ref{fig:triad_interactions_3DPF} in \ref{sup:Primary energetic modes of the periodic flow}), the resulting interactions which appear in the model~\eqref{eq:LV_like_model} are dyadic.
For notation, we employ both \(X\) and \(Y\) as the principal variables of our model and as subscripts to denote quantities associated with these variables.

An extensive parameter scan of the two-equation model shows that the model always converges to a steady solution, and we do not observe an SPO or QCB.
In fact, linear stability analysis of the fixed points of~\eqref{eq:LV_like_model} shows that there are only stable steady solutions (See~\ref{sup:Linear stability analysis of the two-equation model}).
Thus, retaining only this simple interaction between the forced and most energetic modes seems insufficient to reproduce QCB via supercritical bifurcations.
Results of the parameter scan further suggest that the subcritical route to QCB is not present either.

The additional ingredient for QCB turns out to be a small-scale representative and its associated triad interaction terms.
Figure~\ref{fig:three_scales_schematic}~(a) shows the time series of energy \(E(t)\) and energy dissipation rate \(\epsilon(t)\) in the SPO along with partial energy \(E_{X+Y} \equiv \expval{\abs{\vb*{u}_X}^2}/2 + \expval{\abs{\vb*{u}_Y}^2}/2\) and partial energy dissipation rate \(\epsilon_{X+Y} \equiv \nu \qty(\expval{\abs{\vb*{\omega}_X}^2} + \expval{\abs{\vb*{\omega}_Y}^2})\) contained by the forced and primary modes.
While the energy is almost entirely contained in \(E_{X+Y}\), there is a visible difference between the full and partial energy dissipation rates.
This reveals that the rest of the Fourier modes contribute significantly to the dynamics of the energy dissipation, representing the small scales.
We denote the ensemble of these residual modes by \(Z\).
The essential nonlinear interactions of \(Z\) form triads with one mode of the \(Y\)-ensemble and another mode from either the \(Z\)-ensemble or the forced mode \(X\).
These observations lead to a refined three-equation model,
\begin{alignat}{4}
  \dv{X}{t}
    &= - A_1 Y^2 &
    & & &+ A_3 Y Z &
    &- \nu K_X^2 X
    + F, \nonumber \\
  \dv{Y}{t}
    &= + A_1 X Y &
    &- A_2 Z^2 &
    &+ A_4 X Z &
    &- \nu K_Y^2 Y,
    \label{eq:model} \\
  \dv{Z}{t}
    &= &
    &+ A_2 Y Z &
    &- (A_3 + A_4) X Y &
    &- \nu K_Z^2 Z, \nonumber
\end{alignat}
which is represented by a schematic in Fig.~\ref{fig:three_scales_schematic}~(b).
Here, \(A_1,\> A_2 > 0\) and \(A_3,\> A_4 \in \mathbb{R}\) are triad coefficients which retain the discrete Navier--Stokes structure~\eqref{eq:NS_mode}.
We choose the signs and the values of the triad coefficients such that the detailed balance holds in the energy transfer between the three scales.
The signs of \(A_1\) and \(A_2\) are defined so that energy cascades towards small scales: from \(X\) to \(Y\) and \(Y\) to \(Z\).
This two-step energy cascade (for \(A_3 = A_4 = 0\)) is similar to the Obukhov two-stage cascade model~\cite{Obukhov1971_some}.
The triads with coefficients \(A_3\) and \(A_4\) represent the ``non-local'' interactions involving all three scales.
Note that this system is different from, but is of the same level of complexity, as the well-known Lorenz~\cite{Lorenz1963_deterministic} or R\"ossler models~\cite{Rossler1976_equation, Rossler1979_equation}.
An important difference is that each variable denotes a Fourier mode in the Lorenz model, while in our model, it represents a group of modes.

% ++++++++++++++++++++++++++++++++++++++++++++++++++
\subsection{Determination of the parameters}
\label{subsec:Determination of the parameters}
% ++++++++++++++++++++++++++++++++++++++++++++++++++

The model~\eqref{eq:model} is a simplified representation of the SPO, where all Fourier modes are sorted into three scales; the forced mode $X$, the energetic modes $Y$ directly draining energy from $X$ through the $A_1$ interaction, and the small scale modes $Z$ which couple through the local direct cascade interaction $A_2$ with $Y$.
There are also scale non-local interactions represented by $A_3$ and $A_4$.
Even though such a representation of the flow discards details of the actual flow obtained by the DNS, we will fit the model parameters to the DNS data to assess how the model can reproduce actual flow properties.

We can fit six out of eight model constants in~\eqref{eq:model} by comparing them to the DNS of the periodic flow: \(A_i\) with \(i=1,2,3,4\), \(K_\alpha^2\), where \(\alpha \in \qty{X, Y, Z}\), and \(F\).
To do so, we use the energy equations associated with~\eqref{eq:model},
\begin{alignat}{1}
  \dv{E_X}{t}
    &= T_X
    - \epsilon_X
    + P, \nonumber \\
  \dv{E_Y}{t}
    &= T_Y
    - \epsilon_Y,
    \label{eq:model_energy} \\
  \dv{E_Z}{t}
    &= T_Z
    - \epsilon_Z. \nonumber
\end{alignat}
Here, \(E_\alpha \equiv \alpha^2 / 2\) is the energy,
\begin{alignat}{5}
  T_X
    &\equiv -A_1 XY^2 & & &
    &+ A_3 X Y Z, \nonumber \\
  T_Y
    &\equiv +A_1 X Y^2 &
    &- A_2 Y Z^2 &
    &+ A_4 X Y Z,
    \label{eq:model_energy_transfer} \\
  T_Z
    &\equiv &
    &+ A_2 Y Z^2 &
    &- (A_3 + A_4) X Y Z. \nonumber
\end{alignat}
are the energy transfer terms, \(\epsilon_\alpha \equiv 2 \nu K_\alpha^2 E_\alpha\) is the energy dissipation rate, and \(P \equiv FX\) is the energy input rate.
The model parameters are determined by their corresponding quantities of the SPO obtained by DNS.
The resulting values are
\begin{equation}
  A_1 = 0.4,\>
  A_2 = 4,\>
  F = 0.7,\>
  K_X^2 = 2,\>
  K_Y^2 = 5,\> \text{and} \>
  K_Z^2 = 15.
  \label{eq:model_parameters}
\end{equation}
We describe the parameter determination procedure in~\ref{sup:Detailed procedure of the parameter fitting}.
The energy flux coefficients \(A_1\) and \(A_2\) are determined by the energy transfer terms \(T_\alpha\) in~\eqref{eq:model_energy_transfer} while ignoring the nonlocal coefficients \(A_3\) and \(A_4\)~(\ref{eq:evaluate_A1}-\ref{eq:evaluate_A2}).
The forcing coefficient \(F\) is evaluated by \(P\) and the \(X\)-scale energy \(E_X\)~\eqref{eq:evaluate_F}.
The squared characteristic wavenumber \(K_\alpha^2\) is set by \(E_\alpha\) and \(\epsilon_\alpha\) in each scale~\eqref{eq:def_scale_coefficient}.
Note that \(K_X = \sqrt{2}\) of the model parameter can be related to \(\abs{\vb*{k}_f} = \sqrt{2}\) of the forcing~\eqref{eq:body_force} of the DNS.
We remark here that our parameter choice~\eqref{eq:model_parameters} supports the energy cascade picture with \(T_X(t) < 0\): the forced scale \(X\) transfers its energy to smaller scales \((Y, Z)\) on average.
And \(T_Y(t), T_Z(t) > 0\) means that the smaller scales receive energy from the larger scales.
The undetermined parameters of the model are the scale non-local interaction coefficients \(A_3\) and \(A_4\), which can be freely chosen.
The only control parameter is \(\Re \equiv 1/\nu\).
We numerically integrate the model with a fourth-order Runge-Kutta scheme and \(\Delta t = 0.01\) starting from random initial conditions.
See Ref.~\cite{Rackauckas2017_differentialequations} for the solver information.
Our numerical simulations seem to indicate that no periodic solutions exist without the complete non-local interactions: \(A_3 = 0\), \(A_4 = 0\), or \(A_3 + A_4 = 0\).
Conversely, periodic behaviour is observed for a wide range of values when \(A_3 \ne 0\), \(A_4 \ne 0\), and \(A_3 + A_4 \ne 0\).
This observation emphasises the importance of non-local triad interactions for periodic behaviour.

% ++++++++++++++++++++++++++++++++++++++++++++++++++
\subsection{Comparison between the model and the DNS result}
\label{subsec:Comparison between the model and the DNS result}
% ++++++++++++++++++++++++++++++++++++++++++++++++++

\begin{figure}
  \centering
  \includegraphics[width=0.6\textwidth]{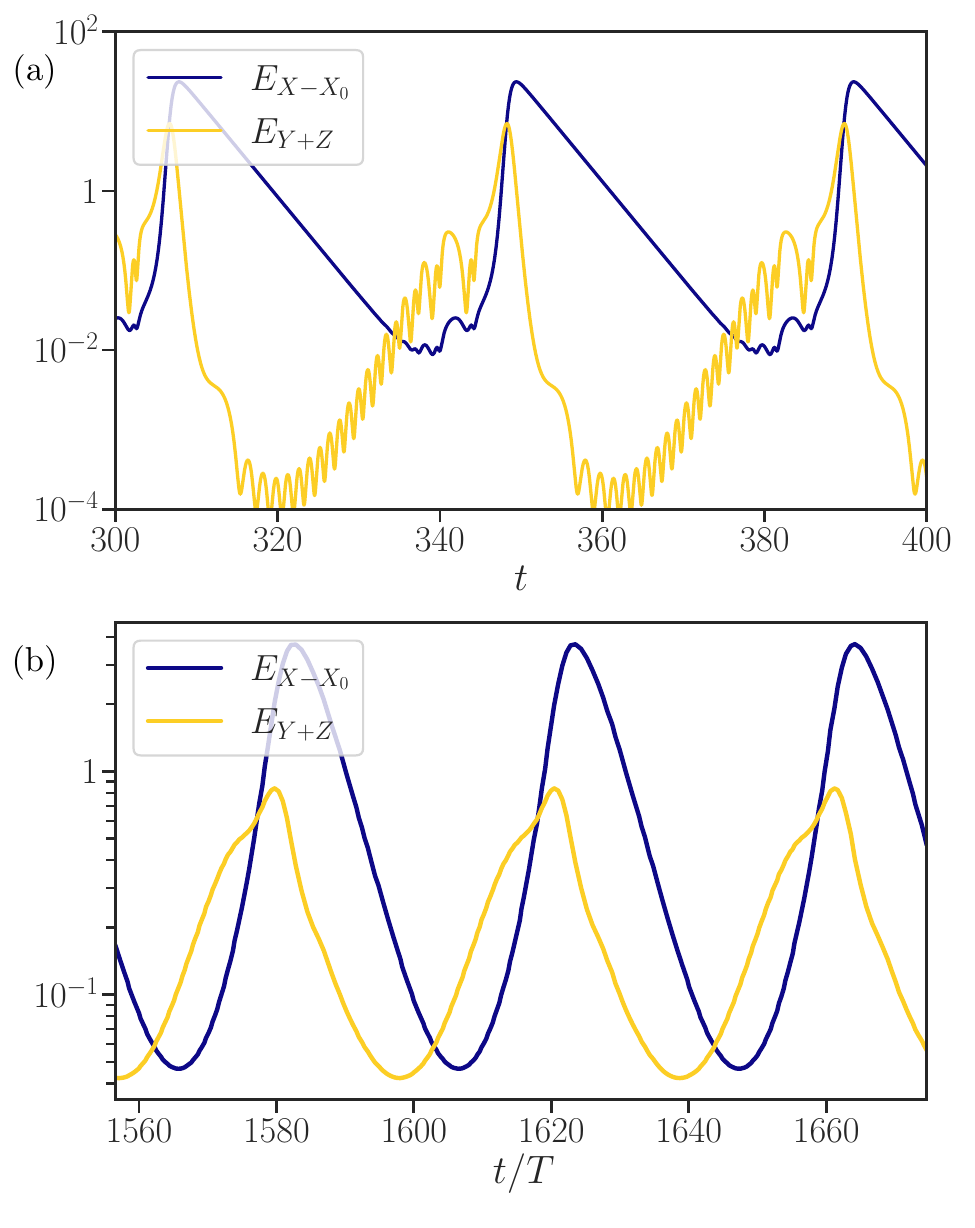}
  \caption{
    Time series of fluctuating energy \(E_{X-X_0}(t)\) of the forced scale and residual energy \(E_{Y+Z}(t)\) of periodic solutions of (a) model~\eqref{eq:model} at \(\Re = 14.05\) and (b) the Navier--Stokes equations~\eqref{eq:Navier--Stokes} at \(\Re = 5.83\).
    Parameters of the model are~\eqref{eq:model_parameters} and \((A_3, A_4) = (0.5, -0.95)\).
    Note that time in panel~(b) is normalised by \(T\).
  }
  \label{fig:timeseries_model_DNS}
\end{figure}

Figure~\ref{fig:timeseries_model_DNS} compares the SPO obtained by the model and the DNS.
Figure~\ref{fig:timeseries_model_DNS}~(a) shows the time series of the model with the parameters~\eqref{eq:model_parameters} and \((A_3, A_4) = (0.5, -0.95)\).
Since the definitions of \(\Re\) are different in the model and DNS, we have chosen a Reynolds number in the model, which allows qualitatively reproducing the DNS results.
We compute two quantities.
One is \(E_{X-X_0} \equiv \qty(X - X_0)^2 / 2\), which is the fluctuating energy of the forced mode around the laminar base flow \(X_0 \equiv F \Re / K_X^2\).
The other quantity \(E_{Y+Z} \equiv Y^2/2 + Z^2/2\) is the energy of the rest of the modes.
We compare them to the corresponding quantities in the DNS of the SPO [Fig.~\ref{fig:timeseries_model_DNS}~(b)], where the base flow is \(\vb*{u}_0 \equiv \vb*{f} / 2 \nu \abs{\vb*{k}_f}^2\), the forced-mode fluctuating energy is \(E_{X-X_0} \equiv \expval{\abs{\vb*{u}_X - \vb*{u}_0}^2}/2\), and \(E_{Y+Z}\) is defined by the energy possessed by the non-forced modes.
We can observe similar periodic behaviour of \(E_{X-X_0}\) and \(E_{Y+Z}\) in the model~\eqref{eq:model} and in the SPO driven by the steady forcing~\eqref{eq:body_force}.
In particular, there are predator-prey-like exponential growth and decay in both systems.
Although fast oscillations are observed in the model but not in the DNS, a close analysis (See~\ref{sup:Primary energetic modes of the periodic flow}) of the DNS of the SPO reveals the presence of rapid oscillations in specific Fourier modes.
These oscillations are compensated by modes that display the same energy oscillations with an opposite phase and do not appear in Fig.~\ref{fig:timeseries_model_DNS}~(b).
We stress that this SPO is independent of the exact amplitude of the initial conditions because the present model is a dissipative system.
% This is in contrast to the standard two-species Lotka--Volterra equations.
% In such conservative systems, the initial condition sets the exact amplitude of the orbit in two-dimensional phase space.

% ==================================================
\section{Dynamics of the model}
\label{sec:Dynamics of the model}
% ==================================================

% ++++++++++++++++++++++++++++++++++++++++++++++++++
\subsection{Bifurcation from SPO to chaos with QCB}
\label{subsec:Bifurcation from SPO to chaos with QCB}
% ++++++++++++++++++++++++++++++++++++++++++++++++++

\begin{figure}
  \centering
  \includegraphics[width=0.6\textwidth]{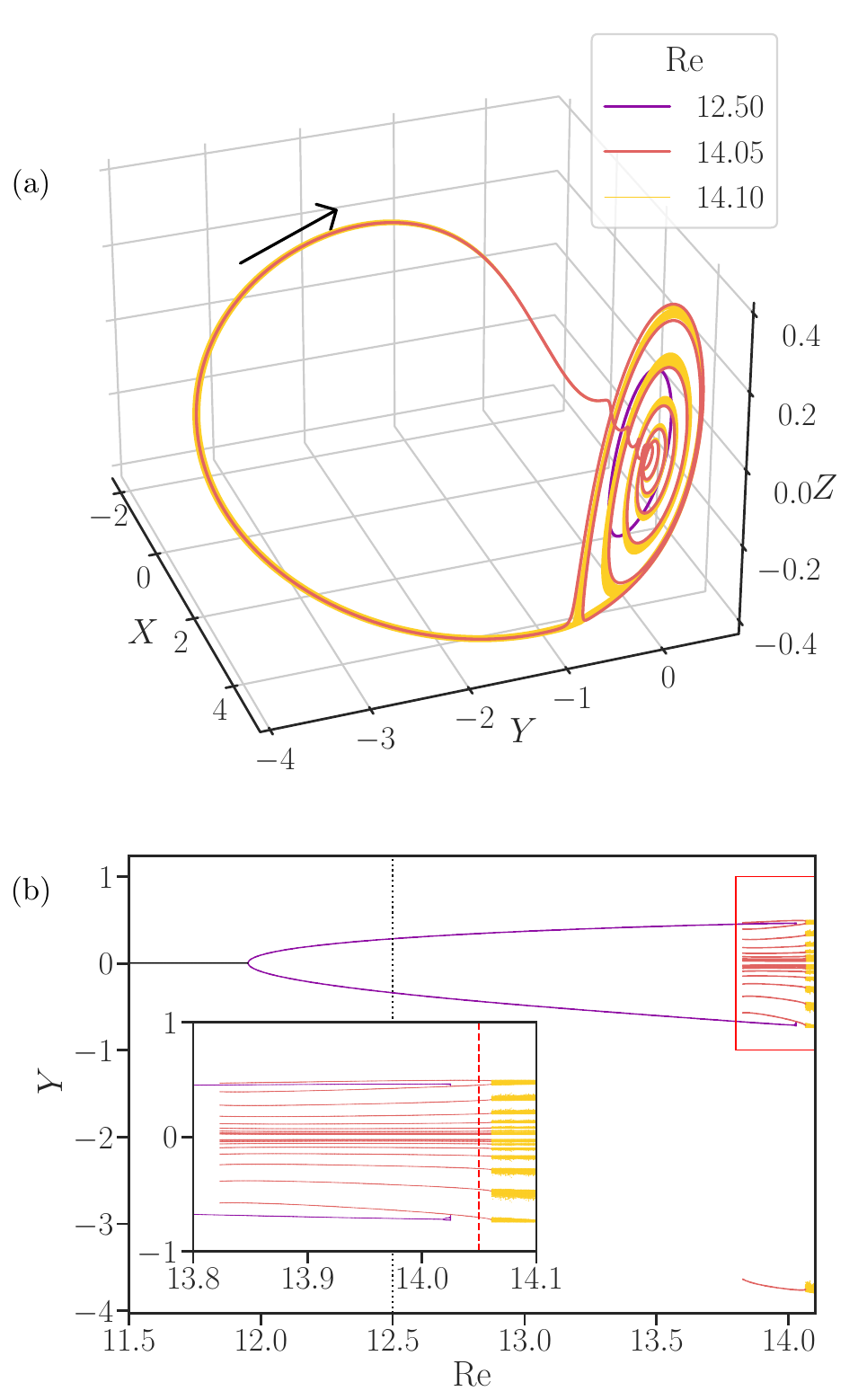}
  \caption{
    (a) Simple periodic (\(\Re = 12.50\)), complex periodic (\(\Re = 14.05\)) [Fig.~\ref{fig:timeseries_model_DNS}~(a)], and chaotic (\(\Re = 14.1\)) orbits of the model~\eqref{eq:model}.
    The parameters are the same as in Fig.~\ref{fig:timeseries_model_DNS}~(a).
    The chaotic orbit is tracked over 100 periods.
    The arrow indicates the direction of the orbit.
    (b) Bifurcation diagram of the model with changing \(\Re\) for the same parameters as in Fig.~\ref{fig:timeseries_model_DNS}~(a).
    We plot the local extrema of \(Y\).
    We have determined the periodicity by Poincar\'{e} analysis.
    The black vertical dotted line corresponds to \(\Re = 12.50\) used in panel (a).
    Inset: close-up in the range shown by the red rectangle in the main plot.
    The red vertical dashed line corresponds to \(\Re = 14.05\) used for panel (a) and Fig.~\ref{fig:timeseries_model_DNS}~(a).
    For both panels, black, purple, orange, and yellow data denote steady, simple periodic, complex periodic, and chaotic solutions, respectively.
  }
  \label{fig:model_orbits_supercritical_bifurcation}
\end{figure}

We observe a chaotic state of the model by varying \(\Re\) from 14.05 to 14.1 while keeping the model parameters as in Fig.~\ref{fig:timeseries_model_DNS}~(a).
Figure~\ref{fig:model_orbits_supercritical_bifurcation}~(a) shows the orbits in phase space for both the periodic (at \(\Re = 14.05\)) and chaotic (at \(\Re = 14.1\)) cases.
The chaotic solution remains close to the SPO as it shows chaotic QCB and is permanent as in the turbulence investigated in \S~\ref{sec:Observation of Quasi-cyclic behaviour}.
Thus, the same model reproduces SPO and chaotic QCB.
Incidentally, the SPO resembles a Shilnikov homoclinic orbit~\cite{Shilnikov1965_a_case}.
We also plot a simpler periodic orbit at \(\Re = 12.5\) in this figure.
It is almost two-dimensional as opposed to the complex three-dimensional periodic and chaotic orbits, suggesting a possible connection with a two-dimensional periodic orbit in the same forcing configuration~\cite[Fig.~4]{vanVeen2018_transitions}.
However, we do not focus on this orbit as it is not directly connected to a chaotic one.

To further assess the behaviour of the system, we draw the bifurcation diagram in Fig.~\ref{fig:model_orbits_supercritical_bifurcation}~(b) with the same parameter set as in Fig.~\ref{fig:timeseries_model_DNS}~(a) and Fig.~\ref{fig:model_orbits_supercritical_bifurcation}~(a).
We observe a supercritical transition from periodic to chaotic solutions at a critical Reynolds number \(\Re_\mathrm{cr} \in \qty[14.060, 14.061]\) [inset of Fig.~\ref{fig:model_orbits_supercritical_bifurcation}~(b)], and, as observed in Fig.~\ref{fig:model_orbits_supercritical_bifurcation}~(a), the chaotic orbit remains close to the SPO.
We note that the solution becomes periodic again when we further increase \(\Re\) beyond the range of Fig.~\ref{fig:model_orbits_supercritical_bifurcation}~(b), probably because the model contains only a small number of degrees of freedom.
The inset of Fig.~\ref{fig:model_orbits_supercritical_bifurcation}~(b) shows that there is a hysteresis in the range \(\Re \in \qty[13.82, 14.03]\), below \(\Re_\mathrm{cr}\), which corresponds to a subcritical bifurcation from a periodic solution to another periodic solution shown in Fig.~\ref{fig:timeseries_model_DNS}~(a).
The appearance of the multiplicity of local extrema corresponds to the spiralling behaviour of the orbit in phase space.
Thus, although the bifurcations from the trivial steady solution to the SPO are rather complicated, that from the SPO to chaos with QCB is simple.
Although there is no clear scenario for the route to turbulence with QCB, the present model results may give us a hint to describe the route in real turbulence.

% ++++++++++++++++++++++++++++++++++++++++++++++++++
\subsection{Subcritical bifurcation to chaos}
\label{subsec:Subcritical bifurcation to chaos}
% ++++++++++++++++++++++++++++++++++++++++++++++++++

\begin{figure}
  \centering
  \includegraphics[width=0.6\textwidth]{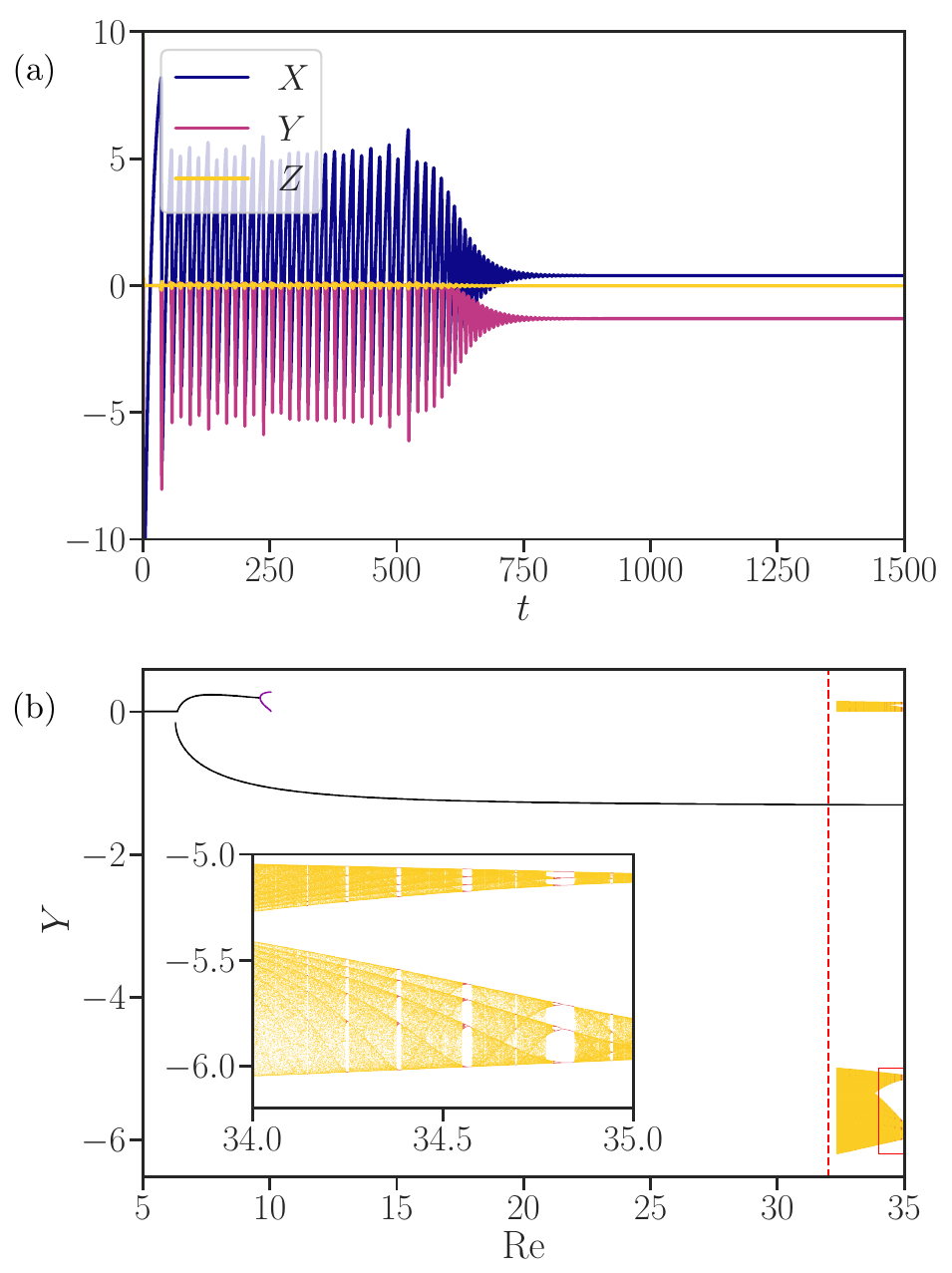}
  \caption{
    (a) Time series of \((X, Y, Z)\) of the model~\eqref{eq:model} with parameters~\eqref{eq:model_parameters} and \((A_3, A_4) = (0.4, -0.5)\) at \(\Re = 32\).
    A random initial condition is used.
    (b) The bifurcation diagram for the same parameter set.
    The red vertical dashed line denotes \(\Re = 32\), which is used for Fig.~\ref{fig:model_timeseries_subcritical_bifurcation}~(a).
    Inset: close-up of the diagram in the range shown by the red rectangle in the main plot.
  }
  \label{fig:model_timeseries_subcritical_bifurcation}
\end{figure}

Since it is well known that, in some cases, turbulence appears via a subcritical transition, here we demonstrate that our model also expresses such a route to chaos.
We stress that we cannot use the strategy above to determine the model parameters since there is no SPO in such a system.
Instead, by varying the undetermined parameters of the model, we observe transient chaos at \((A_3, A_4) = (0.4, -0.5)\) as shown in Fig.~\ref{fig:model_timeseries_subcritical_bifurcation}~(a).
The corresponding bifurcation diagram in Fig.~\ref{fig:model_timeseries_subcritical_bifurcation}~(b) shows a subcritical bifurcation between steady and chaotic solutions around \(\Re \approx 32.3\).
There are bi-stable states for \(33 \lesssim \Re (\lesssim 35)\) of steady and chaotic solutions.
The inset of Fig.~\ref{fig:model_timeseries_subcritical_bifurcation}~(b) shows that there are multiple windows of periodic solutions in the chaotic regime, probably due to the limited number of degrees of freedom of the model~\eqref{eq:model}.

\begin{figure}
  \centering
  \includegraphics[width=0.6\textwidth]{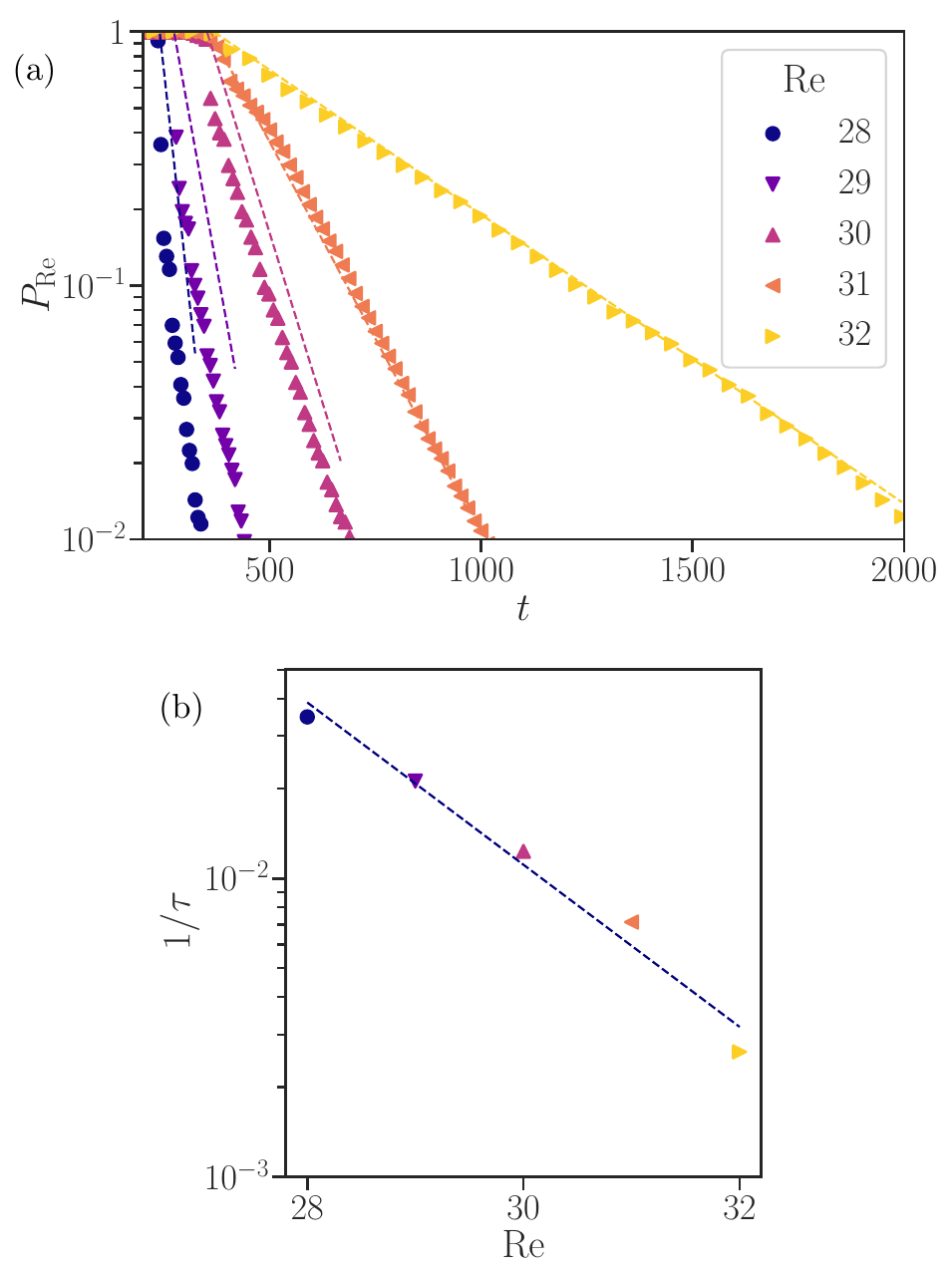}
  \caption{
    (a) Survival probability \(P_{\Re}(t)\) of the transient chaos of the model~\eqref{eq:model} evaluated from 10,000 samples for each \(\Re\).
    The parameter set is the same as in Fig.~\ref{fig:model_timeseries_subcritical_bifurcation}.
    Dashed line denotes exponential fitting by~\eqref{eq:survival_probability} using \(0.01 \le P_{\Re}(t) \le 0.9\).
    (b) The escape rate \(1/\tau\) as a function of \(\Re\).
    The dashed line denotes the exponential fitting by~\eqref{eq:characteristic_time_scale}.
  }
  \label{fig:survival_probability_with_tau_scaling}
\end{figure}

The transient behaviour in Fig.~\ref{fig:model_timeseries_subcritical_bifurcation}~(a) reminds us of the sudden relaminarisation observed in a linearly forced turbulence~\cite{Linkmann2015_sudden}, turbulent Kolmogorov flow~\cite{vanVeen2016_sub}, pipe flow~\cite{Hof2006_finite}, and even in the Lorenz system~\cite{Yorke1979_metastable, Maslennikov2013_dynamic}.
We evaluate the survival probability \(P_{\Re}(t)\), representing how likely the solution remains in a chaotic regime at a given time \(t\), to investigate this phenomenon.
To evaluate \(P_{\Re}(t)\), we identify the relaminarisation time \(t_\mathrm{r}\) by the first time when the local maxima of oscillating energy \(E_{Y-Y_0} \equiv \qty(Y - Y_0)^2 / 2\) becomes smaller than a threshold \(\delta = 1 \times 10^{-3}\).
Here, \(Y_0\) is the stable and steady solution.
Then, the probability \(P_{\Re}(t)\) for given \(t\) can be evaluated by the ratio of a number of samples with \(t_\mathrm{r} < t\) against the number of the whole sample.
We plot \(P_{\Re}(t)\) in Fig.~\ref{fig:survival_probability_with_tau_scaling}~(a) to find that an exponential scaling,
\begin{equation}
  P_{\Re}(t)
    \propto \exp[-\frac{t}{\tau(\Re)}],
  \label{eq:survival_probability}
\end{equation}
fits the data.
The characteristic time scale \(\tau\) in Fig.~\ref{fig:survival_probability_with_tau_scaling}~(b) also displays an exponential scaling,
\begin{equation}
  \tau(\Re)
    \propto \exp[a \Re],
  \label{eq:characteristic_time_scale}
\end{equation}
against \(\Re\).
Although the scaling~\eqref{eq:survival_probability} of \(P_{\Re}(t)\) is consistent with the observations in the previous studies~\cite{Linkmann2015_sudden}, the exponential scaling~\eqref{eq:characteristic_time_scale} of \(\tau(\Re)\) differs from a super-exponential behaviour observed in Ref.~\cite{Linkmann2015_sudden}.
This qualitative difference may also be caused by the minimal number of degrees of freedom in the model.

Note that the Taylor--Green forcing~\eqref{eq:body_force} in the DNS does not permit such a transition since the laminar base flow \(\vb*{u}_0 \equiv \vb*{f} / 2 \nu \abs{\vb*{k}_f}^2\) is linearly unstable.
However, the steady Kolmogorov forcing with a linearly stable laminar base flow exhibits sudden relaminarisations~\cite{vanVeen2016_sub}.
Thus, we can speculate that the model can reflect different forcing set-ups applied to the Navier--Stokes equations by varying the parameters \((A_3, A_4)\).

% ==================================================
\section{Conclusion}
\label{sec:Conclusion}
% ==================================================

The present investigation attempts to construct a minimal model of turbulence with quasi-cyclic behaviour (QCB) in a steady-force driven flow while keeping the structure of the Navier--Stokes equations.
First, through the DNS of Navier--Stokes turbulence, we show that QCB in high-\(\Re\) turbulence is continuously connected to an SPO at small \(\Re\) by extracting the intrinsic periodicity of QCB via a phase averaging technique~(\S~\ref{sec:Observation of Quasi-cyclic behaviour}).
Next, we conduct a mode-by-mode analysis of the SPO to identify the flow's forced, primary energetic, and secondary scales.
We propose the three-equation model~\eqref{eq:model} describing the evolution of such three distinct scales~(\S~\ref{subsec:Construction of the model}).
By adjusting the model parameters, we observe that the model reproduces an SPO similar to that of the DNS~(\S~\ref{subsec:Determination of the parameters}).
We emphasise that scale non-local nonlinear interactions (interactions involving three separate scales) are mandatory for reproducing these dynamics.
Then, we conduct a bifurcation analysis to show that the model also exhibits chaotic QCB via a supercritical bifurcation, which is continuously connected to the SPO~(\S~\ref{subsec:Bifurcation from SPO to chaos with QCB}).
Thus, we conclude that the proposed model reproduces turbulent QCB and its relation to an SPO using a minimum number of degrees of freedom.

Further analysis of the model by varying the undetermined parameters yields transient chaos with sudden relaminarisation, which is also observed in turbulent flow with different forcing set-ups~(\S~\ref{subsec:Subcritical bifurcation to chaos}).
Thus, we speculate that the present model can be a minimal model for certain features of turbulence.

An outstanding open question is how QCB survives in spatially extended flows.
How will the global dynamics change when the forcing is applied to scales smaller than the domain size?
In other words, how will the modes larger than the forced scale alter QCB turbulence, and how can we model it?
Investigating the relation between space and scale locality and temporal dynamics of turbulence is left for further research.

% ==================================================
% Acknowledgments
% ==================================================

\ack{
  All DNS were carried out using the facilities of the PMCS2I (\'Ecole Centrale de Lyon).
  R.A. is supported by the Takenaka Scholarship Foundation, and S.G. is funded by the JSPS Grant-in-Aid for Scientific Research 20H02068.
  The authors thank Dr.~Y.~Duguet for his valuable comments and suggestions on the preprint.
  The authors also thank the anonymous referees for many valuable suggestions to improve the manuscript.
  R.A. deeply thanks Dr.~T.~Tanogami for his insightful comments and discussions and Dr.~T.~Yasuda for his helpful comments on an early draft.
  For the purpose of Open Access, a CC-BY public copyright licence
  has been applied by the authors to the present document and will
  be applied to all subsequent versions up to the Author Accepted
  Manuscript arising from this submission.
}

% ==================================================
\section*{References}
% ==================================================

% \bibliography{biblio}

\providecommand{\newblock}{}

% ==================================================
\clearpage
\appendix
\section*{Appendix}
\section{Direct Numerical Simulations}
\label{sup:Direct Numerical Simulations}
% ==================================================

This appendix describes the detail of the DNS condition.
We use an in-house parallelised code~\cite{Delache2014_scale} to conduct DNS.
It employs a pseudo-spectral method with the 2/3 dealiasing rule for spatial discretisation and the Adams--Bashforth scheme in the time domain.
The initial condition is generated in Fourier space by Rogallo's method~\cite{Rogallo1981_numerical}.

We perform DNS in a \((2\pi)^3\) triply periodic box.
We focus on two distinct flows: three-dimensional periodic and turbulent.
The SPO is obtained by the DNS with \(64^3\) Fourier modes by adjusting the viscosity to \(\nu = 0.102\).
This corresponds to the value of the Reynolds number~\eqref{eq:def_Re_T} of \(\Re = 5.83\).
We use \(128^3\) Fourier modes to simulate turbulent flow at \(\nu = 0.02\) (\(\Re = 29.7\)).
For the phase averaging procedure (See~\ref{sup:Detailed procedure of the phase averaging}), we use the time series in the interval \(1.28 \times 10^3 \le t \le 1.04 \times 10^4\) to guarantee statistical convergence.
Note that we discard the transient part from the analysis.
This interval is approximately \(1.08 \times 10^4 T\) with \(T\) defined in~\eqref{eq:def_Re_T}.

% ==================================================
\section{Detailed procedure of the phase averaging}
\label{sup:Detailed procedure of the phase averaging}
% ==================================================

\begin{figure}
  \centering
  \includegraphics[width=0.6\textwidth]{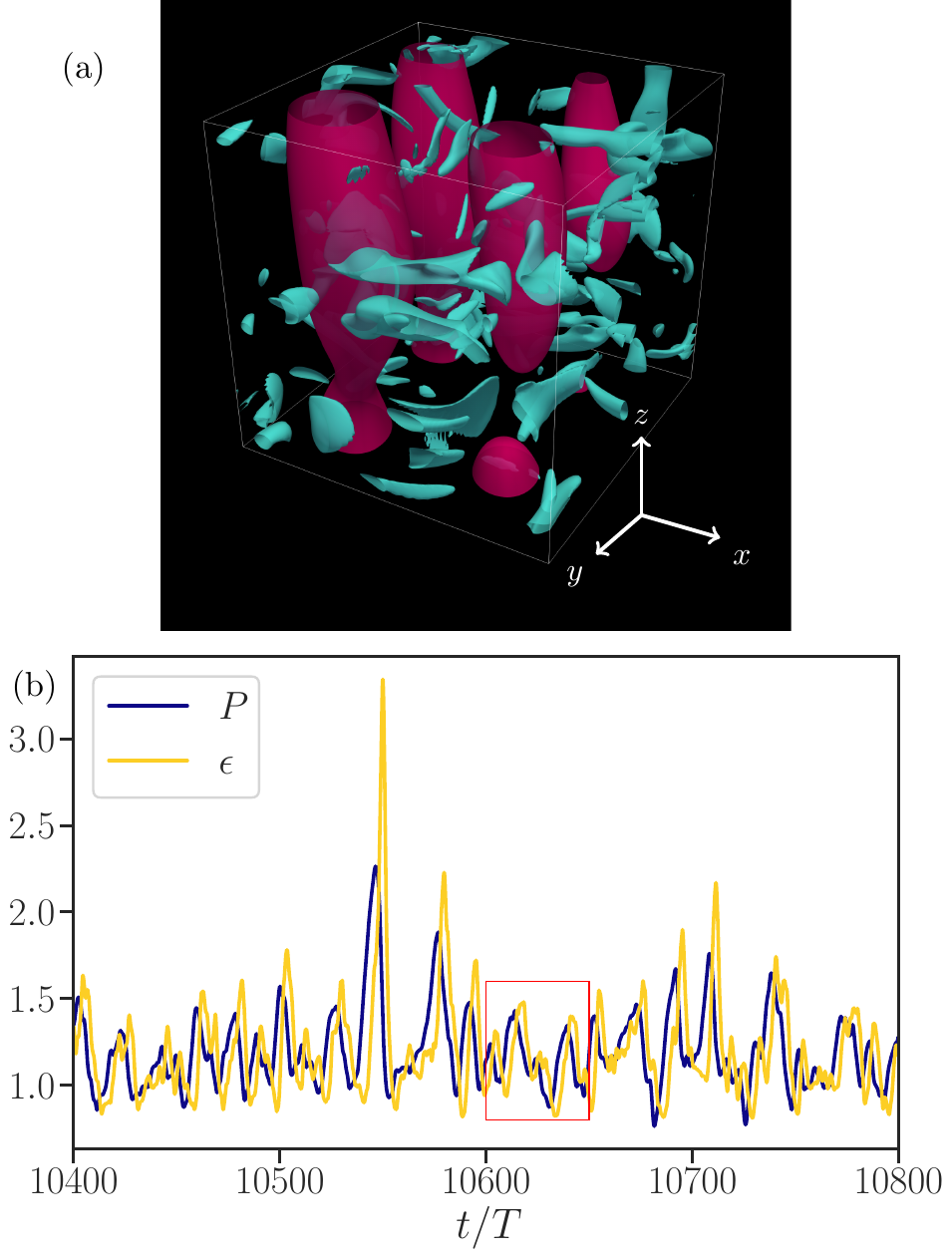}
  \caption{
    (a) A snapshot of the turbulent flow at \(\Re = 29.7\) (\(\Re_\lambda \approx 90\)).
    Isosurfaces of \(\abs{\vb*{\omega}} = 20\) (blue) and low-pass filtered \(\abs{\vb*{\omega}^<} = 4\) (red) are visualised.
    (b) Time series of energy input rate \(P(t)\) and energy dissipation rate \(\epsilon(t)\) of the same flow.
    The red rectangle denotes the time interval examined in Fig.~\ref{fig:QCB_turbulence}~(a).
  }
  \label{fig:turbulence_snapshot_timeseries}
\end{figure}

This appendix explains the detailed procedure of the phase averaging shown in Fig.~\ref{fig:QCB_turbulence}~(b).
We first describe the flow observed at \(\Re = 29.7\) (\(\Re_\lambda\) is about \(90\)) in Fig.~\ref{fig:turbulence_snapshot_timeseries}.
Figure~\ref{fig:turbulence_snapshot_timeseries}~(a) shows isosurfaces of \(\abs{\vb*{\omega}}\) capturing small-scale structures, whereas the forcing-induced columnar vortices emerge by visualising the isosurfaces of \(\abs{\vb*{\omega}^<}\).
Here, \(\vb*{\omega}^< \equiv \grad \times \vb*{u}^<\), which is obtained by applying a low-pass filter to the velocity field, defined as \(\vb*{u}^<(\vb*{x}) \equiv \int \dd{\vb*{r}} G(r/r_0) \vb*{u}(\vb*{x}+\vb*{r})\) with \(G\) being the Gaussian function.
We set \(r_0 = 2\).
Figure~\ref{fig:turbulence_snapshot_timeseries}~(b) shows the temporal evolution of the energy input rate \(P(t)\) and the energy dissipation rate \(\epsilon(t)\).
Both time signals exhibit significant fluctuations with a clear time-delayed correlation.
See Fig.~\ref{fig:QCB_turbulence}~(a) for the 2D projection of the same time series.

\begin{figure}
  \centering
  \includegraphics[width=0.6\textwidth]{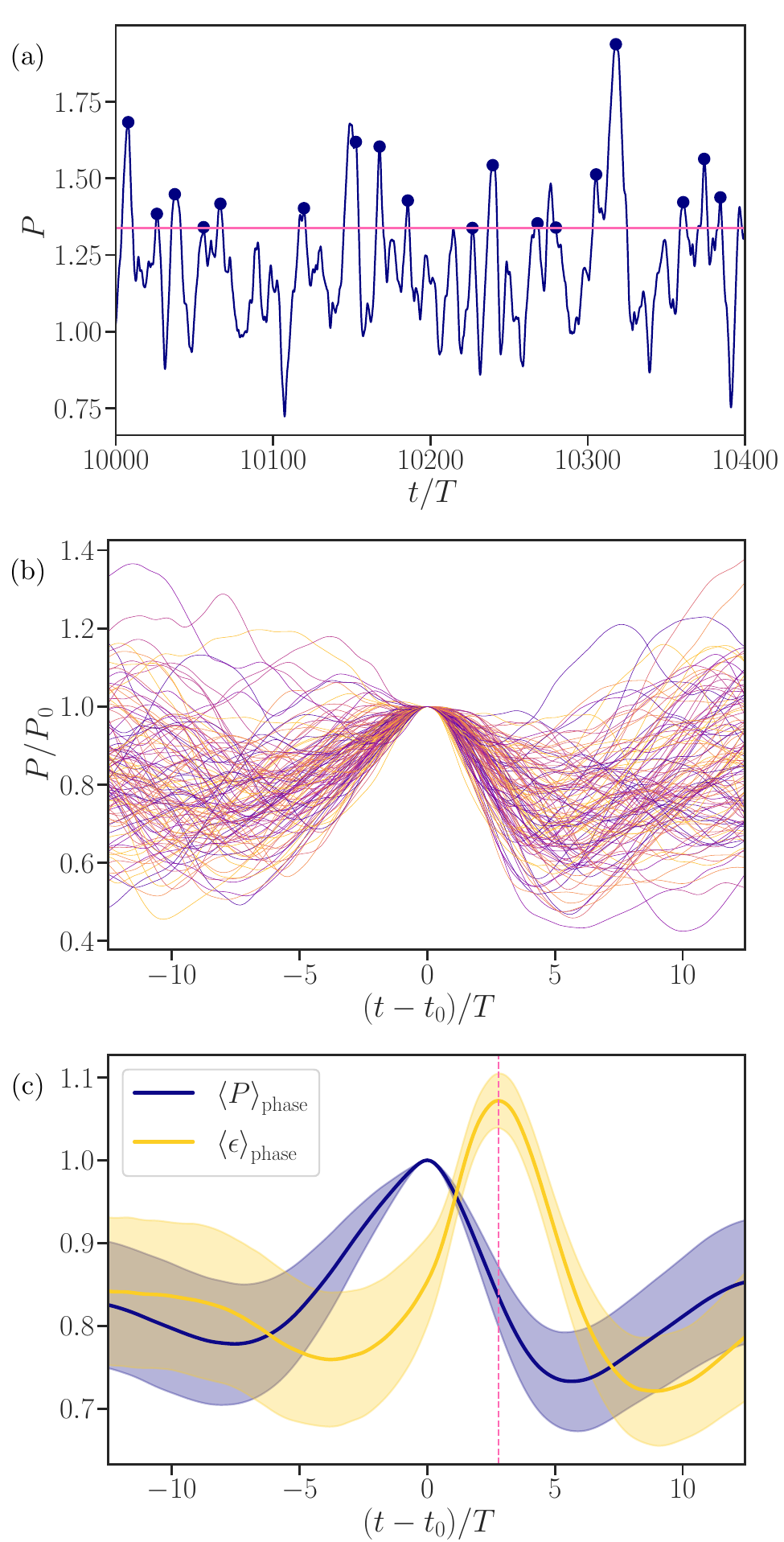}
  \caption{
    (a) Time series of \(P(t) \) with its local maxima \(P_0\) denoted by dots.
    The pink horizontal line corresponds to the threshold of the magnitude \(\expval{P}_t + \sigma(P)\).
    (b) Overlapped segments of the time series of \(P(t)\) around \(t_0\) and normalised by \(P_0\).
    (c) Phase averaged time series of \(P(t)\) and \(\epsilon(t)\).
    Shaded region represents \(\expval{f}_t \pm \sigma(f)\) where \(f\) is \(P(t)\) or \(\epsilon(t)\).
    The pink vertical dashed line indicates the maximum of \(\expval{\epsilon}_\mathrm{phase}\), denoting the average time delay between \(P(t)\) and \(\epsilon(t)\).
  }
  \label{fig:phase_averaging}
\end{figure}

To conduct the phase-averaging, first, we pick up the local maxima of \(P(t)\) [Fig.~\ref{fig:phase_averaging}~(a)] with the following two criteria:
(i) It must be larger than \(\expval{P}_t + \sigma(P)\) where \(\expval{\cdot}_t\) and \(\sigma(\cdot)\) denote the time average and the standard deviation, respectively.
The horizontal pink line indicates this value in Fig.~\ref{fig:phase_averaging}~(a).
(ii) The temporal gap between two consecutive local maxima must be larger than \(\tau_\text{max} / 2\) where \(\tau_\text{max}\) is the time for the second peak of the autocorrelation function of \(P\) (note that the first peak is at \(\tau = 0\)).
We denote the identified local maximum of \(P(t)\) and the corresponding time by \(P_0\) and \(t_0\), respectively.
Second, the segments of the time series of \(P(t)\) around the local maximum \(P_0\) are overlapped, as shown in Fig.~\ref{fig:phase_averaging}~(b).
We normalise the segments by \(P_0\) to avoid overestimation due to huge intermittent peaks.
Third, we compute the average over the overlapped and normalised time series to obtain the phase averaged time series \(\expval{P}_\mathrm{phase}\) shown in Fig.~\ref{fig:phase_averaging}~(c).

We apply a similar procedure to \(\epsilon(t)\).
However, the time is shifted for \(t_0\), and \(\epsilon(t)\) is normalised by \(P_0\) so that we can evaluate the time delay and the relative amplitude difference between the two quantities.
The pink vertical dashed line shows the time delay in Fig.~\ref{fig:phase_averaging}~(c), which is \(2.80 T\).
Figure~\ref{fig:QCB_turbulence}~(b) is a parametric plot of Fig.~\ref{fig:phase_averaging}~(c).

% ==================================================
\section{Primary energetic modes of the periodic flow}
\label{sup:Primary energetic modes of the periodic flow}
% ==================================================

\begin{figure}[t]
  \centering
  \includegraphics[width=\textwidth]{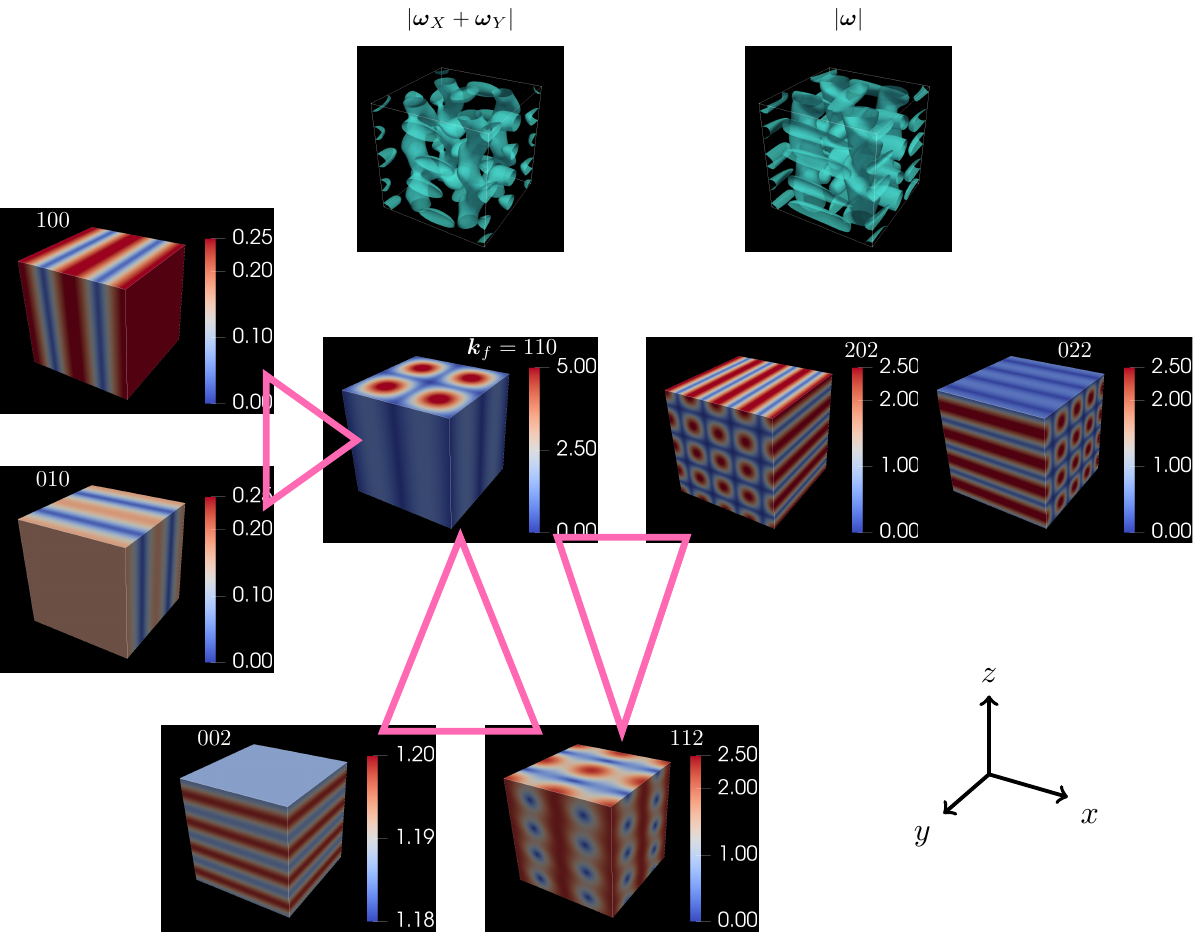}
  \caption{
    Schematic of forced (centre) plus six primary energetic (surrounding) Fourier modes in the SPO at \(\Re = 5.83\).
    Visualisations show distributions of \(\abs{\vb*{\omega}}\) at the same instance.
    The three-digit numbers on the visualisations indicate three components of wavevector \(k_x k_y k_z\).
    Note that all the possible sign combinations \((\pm k_x, \pm k_y, \pm k_z)\) are gathered.
    Triangles denote the possible triad interactions.
    On top, we compare the isosurfaces of \(\abs{\vb*{\omega}_{X} + \vb*{\omega}_{Y}}\) with  \(\abs{\vb*{\omega}}\) of the full flow.
    Here, \(\vb*{\omega}_{X}\) and \(\vb*{\omega}_{Y}\) denote the vorticity of the forced and the primary energetic modes, respectively.
  }
  \label{fig:triad_interactions_3DPF}
\end{figure}

This appendix shows the detailed behaviour of the seven most energetic modes.
Figure~\ref{fig:triad_interactions_3DPF} shows \(\abs{\vb*{\omega}}\) distributions of these seven modes.
Triangles indicate combinations of different modes where energy transfer via triad interactions is possible.
Figure~\ref{fig:triad_interactions_3DPF} also compares isosurfaces of \(\abs{\vb*{\omega}}\) of the sum of these seven primary energetic modes and that of all modes.
We find similar principal structures: the large columnar vortices and the small counter-rotating pairs of vortices.
Here, we denote the velocity field consisting of the forced mode by \(\vb*{u}_X\) and the other six primary energetic modes by \(\vb*{u}_Y\).
The corresponding vorticity fields are denoted by \(\vb*{\omega}_X\) and \(\vb*{\omega}_Y\), respectively.

\begin{figure}
  \centering
  \includegraphics[width=0.7\textwidth]{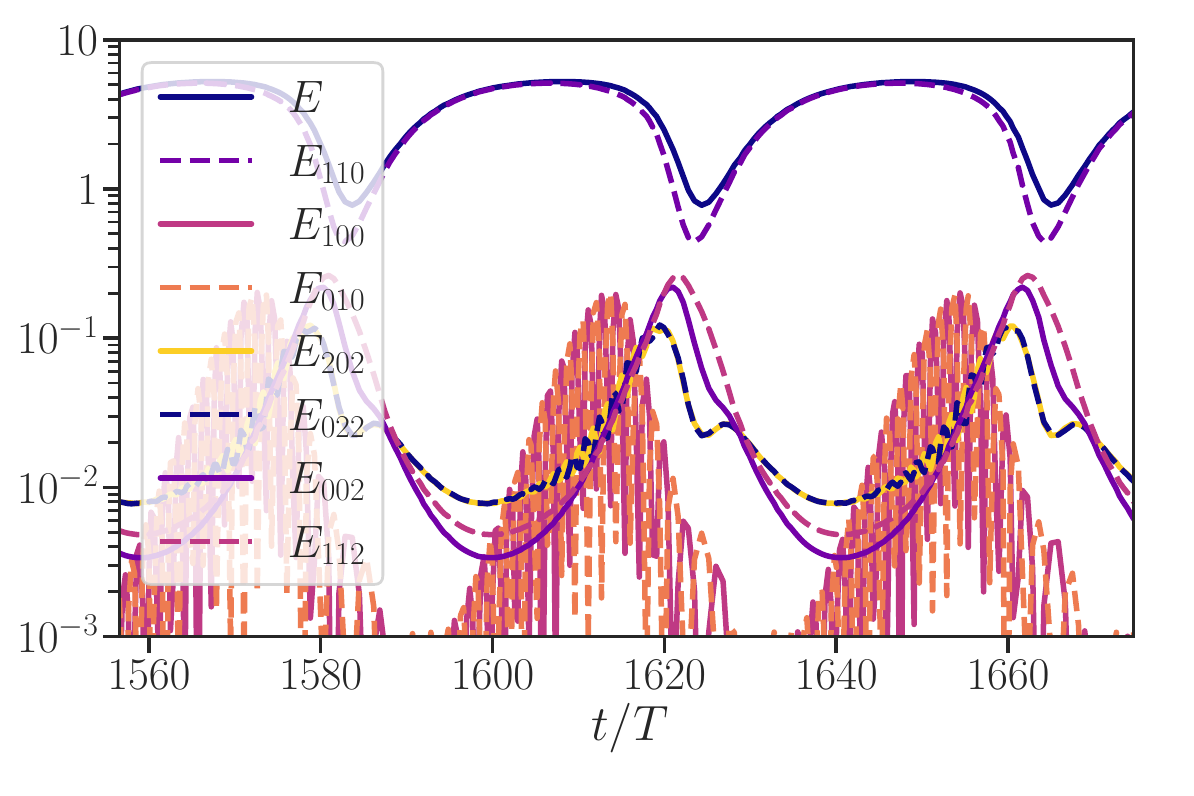}
  \caption{
    Time series of energy of the forced and six primary energetic modes in the SPO.
    \(E_{k_x k_y k_z}\) denotes the energy summed-up for the wavevectors \((\pm k_x, \pm k_y, \pm k_z)\).
    Total energy \(E(t)\) is also shown for reference.
  }
  \label{fig:energy_seven_modes}
\end{figure}

We plot the time series of the energy of the forced and six primary modes of the SPO at \(\Re = 5.83\) in Fig.~\ref{fig:energy_seven_modes}.
Although the energy \(E_{110}(t)\) of the forced mode dominates, which is approximately equal to the total energy \(E(t)\), we observe a distinctive difference between \(E(t)\) and \(E_{110}(t)\) when the primary scale energies are excited.
By summing up the contributions of these seven modes, we obtain \(E_{X+Y}\) shown in Fig.~\ref{fig:three_scales_schematic}~(a).
We also note that there are fast oscillations in \(E_{100}(t)\) and \(E_{010}(t)\).
However, these two modes are compensated with each other, and such rapid dynamics are not visible in \(E_{X+Y}(t)\) [Fig.~\ref{fig:three_scales_schematic}~(a)].
This observation explains why there are no fast oscillations in the time series shown in Fig.~\ref{fig:timeseries_model_DNS}~(b).

% ==================================================
\section{Linear stability analysis of the two-equation model}
\label{sup:Linear stability analysis of the two-equation model}
% ==================================================

In this appendix, we show the results of the linear stability analysis of the fixed points of~\eqref{eq:LV_like_model}.
There are two kinds of fixed points: namely,
\begin{equation}
  \overline{\vb*{X}}_1
    = \qty(\frac{F}{\nu K_X^2}, 0) \qand
  \overline{\vb*{X}}_2
    = \qty(\frac{\nu K_Y^2}{A}, \pm\frac{1}{A} \sqrt{AF - \nu^2 K_X^2 K_Y^2}),
  \label{eq:LV_like_model_fixed_points}
\end{equation}
where \(\overline{\vb*{X}} \equiv (\overline{X}, \overline{Y})\).
Note that the fixed points \(\overline{\vb*{X}}_2\) exist only for \(\nu < \sqrt{AF} / K_X K_Y\).
The perturbation \((x,y)\) in the vicinity of the fixed points \((\overline{X}, \overline{Y})\) obeys
\begin{alignat}{3}
  \dv{x}{t}
      &= - A \qty(\overline{Y}^2 + 2\overline{Y} y) &
      &- \nu K_X^2 \qty(\overline{X} + x)
      + F, \\
    \dv{x}{t}
      &= + A \qty(\overline{X}~\overline{Y} + \overline{X} y + x \overline{Y}) &
      &- \nu K_Y^2 \qty(\overline{Y} + y),
  \label{eq:LV_like_model_linearized}
\end{alignat}
where we have neglected second-order terms \(x^2, y^2\), and \(xy\).
The Jacobian matrix is then expressed as
\begin{equation}
  \vb*{J} = \mqty(
    - \nu K_X^2 & -2A \overline{Y} \\
    A \overline{Y} & A \overline{X} - \nu K_Y^2
  ),
  \label{eq:LV_like_model_Jacobian}
\end{equation}
whose eigenvalues are
\begin{equation}
  \lambda
    = -\frac12 \qty[-A \overline{X} + \nu \qty(K_X^2 + K_Y^2)]
    \pm \frac12 \sqrt{\qty[A \overline{X} + \nu \qty(K_X^2 - K_Y^2)]^2 - 8 A^2 \overline{Y}^2}.
\end{equation}
The eigenvalues for \(\vb*{X}_1\) are
\begin{equation}
  \lambda^{(\vb*{X}_1)}_1
    = - \nu K_X^2, \quad
  \lambda^{(\vb*{X}_1)}_2
    = \frac{AF}{\nu K_X^2} - \nu K_Y^2,
\end{equation}
which are both negative for \(\nu > \sqrt{AF}/K_X K_Y\).
Therefore, \(\overline{\vb*{X}}_1\) is stable for \(\nu > \sqrt{AF}/K_X K_Y\), and a pitchfork bifurcation takes place at \(\nu = \sqrt{AF}/K_X K_Y\).
Then, for \(\nu < \sqrt{AF}/K_X K_Y\), \(\overline{\vb*{X}}_2\) exists, which is stable irrespective of \(\nu\) because the eigenvalues are
\begin{equation}
  \lambda^{(\vb*{X}_2)}_{1,2}
    = -\frac{\nu K_X^2}{2}
    \pm \frac{\sqrt{-8AF + \nu^2 K_X^2 \qty(1 + 8K_Y^2)}}{2}.
\end{equation}

% ==================================================
\section{Detailed procedure of the parameter fitting}
\label{sup:Detailed procedure of the parameter fitting}
% ==================================================

\begin{figure}
  \centering
  \includegraphics[width=0.6\textwidth]{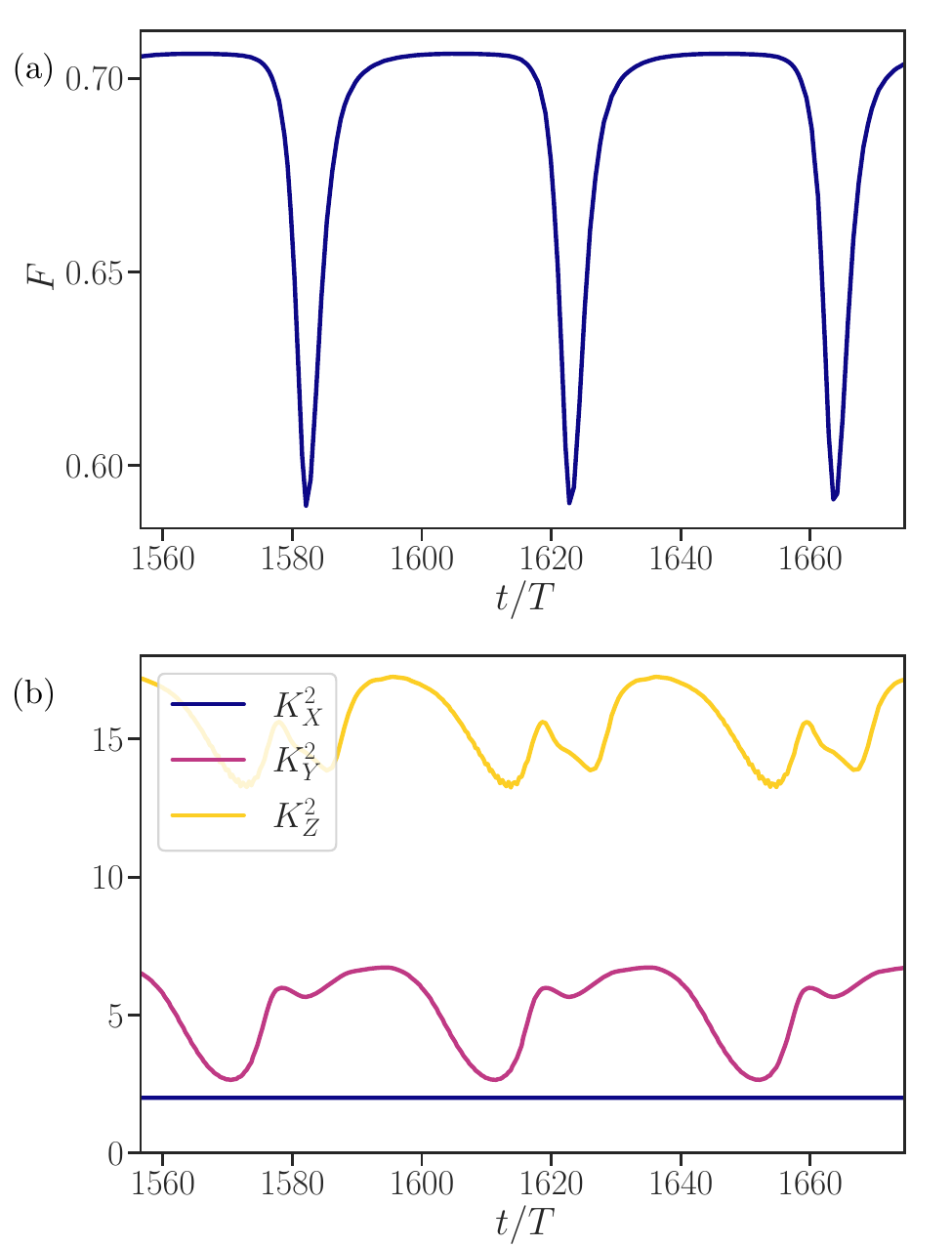}
  \caption{
    Time series of (a) forcing coefficient \(F(t)\) and (b) scale coefficients \(K^2_\alpha(t)\) in the DNS of the SPO.
  }
  \label{fig:F_P_scale_coefficients}
\end{figure}

This appendix discusses the detailed procedure of the parameter fitting~\eqref{eq:model_parameters} of the three-equation model~\eqref{eq:model}.
Figure~\ref{fig:F_P_scale_coefficients}~(a) shows the time evolution of the forcing coefficient defined by
\begin{equation}
  F(t) \equiv \frac{P}{\sqrt{2 E_X}}.
  \label{eq:evaluate_F}
\end{equation}
We estimate the model parameter \(F = 0.7\), since the time average \(\expval{F(t)}_t = 0.696\).
The periodic drops of \(F(t)\) are associated with a phase-desynchronisation between the forcing and the forcing-induced velocity field, \(\vb*{u}_X\).

We also compute the scale factors
\begin{equation}
  K_\alpha^2(t) \equiv \frac{ \epsilon_\alpha }{ 2 \nu E_\alpha } \quad
  (\alpha \in \qty{X, Y, Z}).
  \label{eq:def_scale_coefficient}
\end{equation}
Figure~\ref{fig:F_P_scale_coefficients}~(b) shows their temporal evolutions.
The forced scale factor \(K_X^2(t) = 2\) is constant, since it corresponds to \(\vb*{k}_f = (\pm 1, \pm 1, 0)\) mode.
On the other hand, \(K_Y^2(t)\) and \(K_Z^2(t)\) fluctuate, reflecting the competition of different Fourier modes in these scales.
We estimate the model parameters by \(K_Y^2 = 5\) and \(K_Z^2 = 15\), since \(\expval{K_Y^2 (t)}_t = 4.97\) and \(\expval{K_Z^2 (t)}_t = 15.4\), respectively.

\begin{figure}
  \centering
  \includegraphics[width=0.6\textwidth]{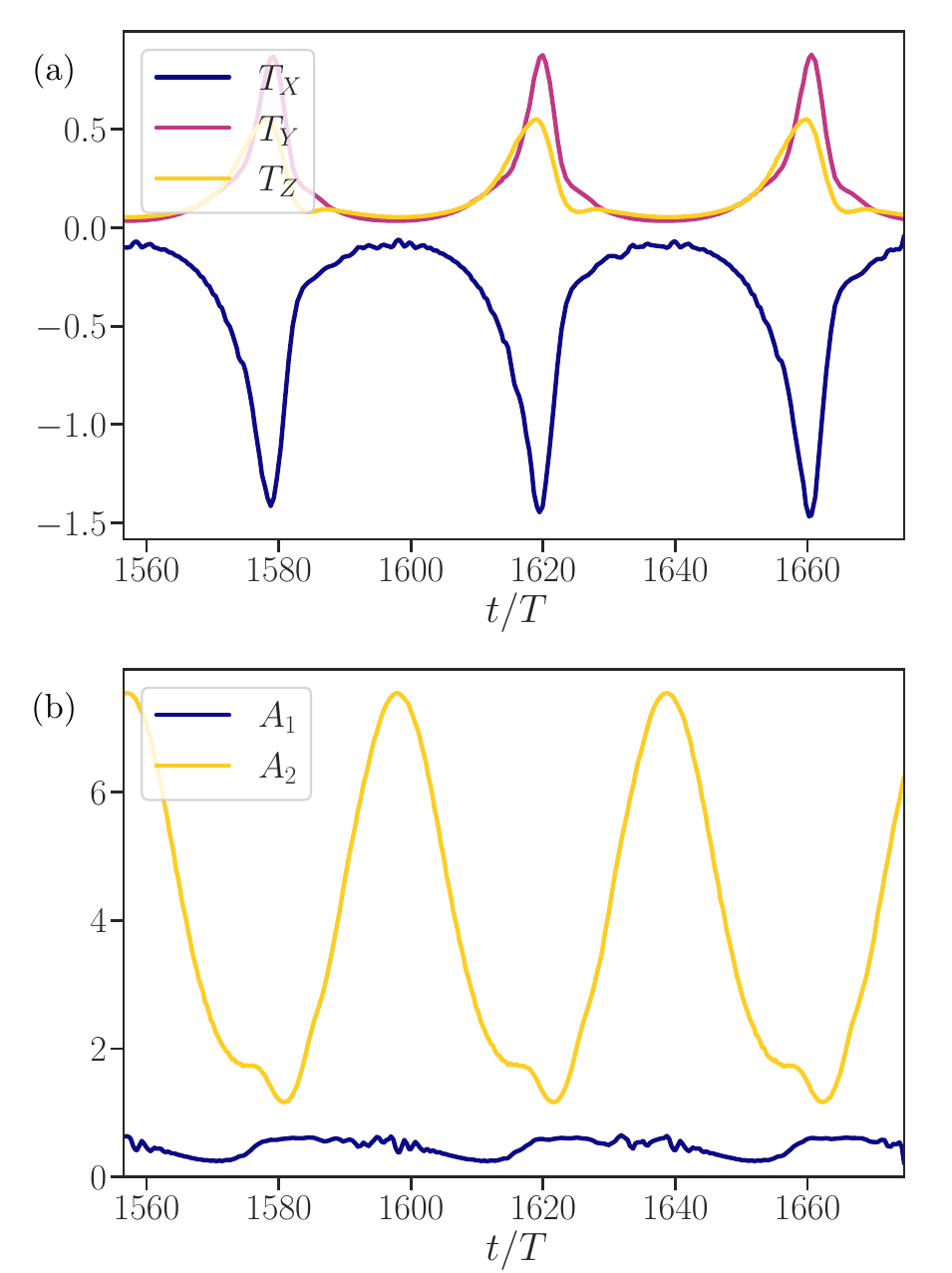}
  \caption{
    Time series of (a) energy transfer terms \(T_\alpha(t)\) and (b) transfer coefficients \(A_i(t)\) in the DNS of the SPO.
  }
  \label{fig:energy_transfer_A1_A2}
\end{figure}

To obtain rough estimates of the scale local coefficients \(A_1\) and \(A_2\), we compute the average energy transfer rate from \(X\) to \(Y\) and \(Y\) to \(Z\) while ignoring the scale non-local interactions by setting \(A_3 = A_4 = 0\).
In this way, the energy transfer terms~\eqref{eq:model_energy_transfer} of the energy equation~\eqref{eq:model_energy} of the model are approximated by
\begin{alignat}{3}
  T_X(t)
    &\approx - A_1 XY^2, & \nonumber \\
  T_Y(t)
    &\approx + A_1 X Y^2 &
    &- A_2 Y Z^2,
    \label{eq:model_energy_transfer_A3=A4=0} \\
  T_Z(t)
    &\approx &
    &+ A_2 Y Z^2. \nonumber
\end{alignat}
Figure~\ref{fig:energy_transfer_A1_A2}~(a) shows their time series by the DNS of the SPO.
\(T_X(t) < 0\) supports the energy cascade picture; the forced scale \(X\) is transferring energy to smaller scales \((Y, Z)\) on average.
Similarly, \(T_Y(t), T_Z(t) > 0\) means that these smaller scales receive energy from the larger scales.
We then evaluate the time-dependent coefficients,
\begin{alignat}{2}
  A_1(t)
    &\approx - \frac{T_X}{X Y^2} &
    &= - \frac{1}{2\sqrt{2}} \frac{T_X}{\sqrt{E_X} E_Y},
    \label{eq:evaluate_A1} \\
  A_2(t)
    &\approx \frac{T_Z}{Y Z^2} &
    &= \frac{1}{2\sqrt{2}} \frac{T_Z}{\sqrt{E_Y} E_Z}.
    \label{eq:evaluate_A2}
\end{alignat}
Again, we neglect the scale non-local interactions \((A_3 = A_4 = 0)\) in these expressions.
The result is shown in Fig.~\ref{fig:energy_transfer_A1_A2}~(b), and we estimate \(A_1 = 0.4\) and \(A_2 = 4\) as the model parameters from the time-averaged values \(\expval{A_1 (t)}_t = 0.440\) and \(\expval{A_1 (t)}_t = 4.04\), respectively.

The above argument allows us to determine the model parameters in~\eqref{eq:model_parameters}.
The non-local interaction coefficients \(A_3\) and \(A_4\) are left to be determined.
In \S~\ref{subsec:Determination of the parameters}, we vary these two parameters to investigate the model properties.

\end{document}